\documentclass[aps,twocolum,amsmath,amssymb,a4paper,allowtoday,accepted=2023-08-18]{quantumarticle}
\pdfoutput=1
\usepackage{amsmath,braket,amssymb}
\usepackage[final]{graphicx}
\usepackage{subfigure}
\usepackage{xcolor}
\usepackage[english]{babel}
\usepackage[utf8]{inputenc}
\usepackage{color}
\usepackage{mathtools}
\usepackage{empheq}
\usepackage{tensor}

\usepackage[colorlinks,citecolor=darkBlue,linkcolor=darkBlue,
urlcolor=blue,hyperindex]{hyperref}

\usepackage[normalem]{ulem}
\normalfont

\graphicspath{{./images/},{./imagesAppendix/}}

\definecolor{forestgreen}{rgb}{0.08, 0.4, 0.13}
\definecolor{darkBlue}{rgb}{0.08, 0.13, 0.4}

\definecolor{THc}{rgb}{0.9,0.3,0.2}

\usepackage{amsthm}
\theoremstyle{definition}

\theoremstyle{plain}

\newcommand{\idg}[1]{{\bfseries #1)}}

\newcommand{\subfigimg}[3][,]{%
	\setbox1=\hbox{\includegraphics[#1]{#3}}% Store image in box
	\leavevmode\rlap{\usebox1}% Print image
	\rlap{\hspace*{2pt}\raisebox{\dimexpr\ht1-0.5\baselineskip}{{\bfseries \large\textsf{#2}}}}% Print label
	\phantom{\usebox1}% Insert appropriate spcing
}

\begin{document}
	
\title{Stabilizer entropies and nonstabilizerness monotones}
	
\author{Tobias Haug}
\email{tobias.haug@u.nus.edu}
\affiliation{QOLS, Blackett Laboratory, Imperial College London SW7 2AZ, UK}

\author{Lorenzo Piroli}
\affiliation{Philippe Meyer Institute, Physics Department, \'{E}cole Normale Sup\'{e}rieure (ENS), Universit\'{e} PSL, 24 rue Lhomond, F-75231 Paris, France}

\date{\today}
	
\begin{abstract}
We study different aspects of the stabilizer entropies (SEs) and compare them against known nonstabilizerness monotones such as the min-relative entropy and the robustness of magic. First, by means of explicit examples, we show that, for R\'enyi index $0\leq n<2$, the SEs are not monotones with respect to stabilizer protocols which include computational-basis measurements, not even when restricting to pure states (while the question remains open for $n\geq 2$). Next, we show that, for any R\'enyi index, the SEs do not satisfy a strong monotonicity condition with respect to computational-basis measurements. We further study SEs in different classes of many-body states. We compare the SEs with other measures, either proving or providing numerical evidence for inequalities between them.
Finally, we discuss exact or efficient tensor-network numerical methods to compute SEs of matrix-product states (MPSs) for large numbers of qubits. In addition to previously developed exact methods to compute the R\'enyi SEs, we also put forward a scheme based on perfect MPS sampling, allowing us to compute efficiently the von Neumann SE for large bond dimensions. 
\end{abstract}
	
\maketitle

\section{Introduction}

Stabilizer states and Clifford operations are a pillar of quantum information theory~\cite{gottesman1997stabilizer,gottesman1998theory,gottesman1998heisenberg,aaronson2004improved,nielsen2011quantum}. For instance, the stabilizer formalism is at the basis of many prototypical quantum error-correcting  codes~\cite{kitaev2003fault,eastin2009restriction}, where Clifford operations can be implemented fault tolerantly~\cite{shor1996fault,preskill1998fault}. A prominent feature of Clifford circuits and stabilizer states is that they can be efficiently simulated on a classical computer~\cite{gottesman1997stabilizer,gottesman1998theory,gottesman1998heisenberg,aaronson2004improved}. While this makes them easy to study, it implies that they have to be supplemented with suitable nonstabilizer ancillary states in order to achieve universal quantum computation~\cite{bravyi2005universal,campbell2017roads}.

It is an  important task to quantify the degree to which a quantum state cannot be prepared by Clifford gates, i.e., how much it deviates from a stabilizer state. This property, known as nonstabilizerness or magic, has received significant attention in the quantum information literature. Similar to quantum entanglement, nonstabilizerness allows for a resource theory~\cite{chitambar2019quantum}, where Clifford operators and stabilizer states are free resources. In this context, different nonstabilizerness monotones have been proposed as measures to quantify it~\cite{campbell2011catalysis,veitch2014resource,howard2017application,wang2019quantifying,beverland2020lower,jiang2021lower, liu2022many,bu2022complexity,bu2019efficient}.

More recently, nonstabilizerness was considered in the context of many-body quantum physics~\cite{white2021conformal,sarkar2020characterization,sewell2022mana,oliviero2022magic,liu2022many}. Here, a major difficulty is that measures of nonstabilizerness are typically very hard to compute (although an efficient measurement protocol for quantum computers has been recently reported~\cite{haug2022scalable}). This is especially true in the prototypical case of qubit systems and, more generally, when the local Hilbert space dimension is even~\cite{campbell2012magic,anwar2014fast,campbell2014enhanced}. 

In order to make progress, useful functions to quantify nonstabilizerness, the so-called stabilizer entropies (SEs),  were introduced in Ref.~\cite{leone2022stabilizer}, cf. also~\cite{leone2023nonstabilizerness,oliviero2022magic,leone2023phase,odavic2022complexity,chen2022magic}. They are expressed in terms of the expectation values of all Pauli strings and, differently from several known monotones, they do not require minimization procedures, making their evaluation simpler. In addition, while the computation of the SE of a typical state is exponentially costly in the system size~\cite{oliviero2022magic}, it can be computed efficiently for the important class of matrix product states (MPSs)~\cite{haug2022quantifying}. This allowed for a detailed study of SEs in  one-dimensional systems~\cite{haug2022quantifying}, substantiating previously suggested connections between nonstabilizerness and criticality~\cite{white2021conformal}. It is also worth mentioning that SEs can be probed experimentally by randomized measurement protocols~\cite{oliviero2022measuring} or Bell measurements~\cite{haug2022scalable,haug2023efficient}. 

The SE of a pure state $\ket{\psi}$ satisfies the following properties~\cite{leone2022stabilizer}: (i) it is zero if and only if $\ket{\psi}$ is a stabilizer state; (ii) it is invariant under Clifford unitaries; (iii) it is additive under tensor product. SEs can be also defined for mixed states. However, if we define the mixed stabilizer states as the convex hull of the pure stabilizer states, then property (i) does not hold anymore. While the latter can be recovered if a restricted definition of mixed stabilizer states is used~\cite{leone2022stabilizer}, in this work we will restrict to the case of pure states. 

From the resource-theory point of view, an important open question  is whether SEs are monotones with respect to the set of stabilizer protocols~\cite{veitch2014resource,heimendahl2022axiomatic}. Together with Clifford unitaries, these include Pauli measurements and the possibility to discard qubits or operations conditioned on measurement outcomes. More generally, it is also natural to ask how the SE compares to known magic monotones, for instance in terms of order relations between them.

In this work we address these questions, making further progress on the characterization of SEs. We first present a series of general results about monotonicity properties of SEs and their relation with known monotones, namely the min-relative entropies~\cite{bravyi2019simulation,liu2022many} and the robustness of magic~\cite{howard2017application,regula2017convex,seddon2021quantifying}. For R\'enyi index $0\leq n<2$, we show that the SE is not a monotone for general stabilizer protocols, including protocols involving only pure states. In addition, we show that the SEs do not satisfy a strong monotonicity condition with respect to computational-basis measurements for any R\'enyi index. 
This result contradicts a previous statement in the literature stating that the R\'enyi-$1/2$ SE is a strong monotone~\cite{hahn2022quantifying}, cf. also~\cite{hahn2023erratum}.

Next, we study SEs in different classes of many-body states and compare them against known magic monotones.  We compare the SEs with other measures, either proving or providing numerical evidence for inequalities between them. Finally, we discuss tensor-network methods for the efficient computations in MPSs, extending the results of Ref.~\cite{haug2022quantifying}. In particular, we introduce a new computational scheme based on perfect MPS sampling~\cite{ferris2012perfect}. This allows us to access the von Neumann stabilizer entropy, which was outside the reach of the exact method developed in Ref.~\cite{haug2022quantifying}.

Finally, it is important to mention that, independent of their monotonicity properties, SREs have already found numerous applications in different areas of quantum information theory. For instance, they allow us to bound the cost of certain quantum-certification~\cite{leone2023nonstabilizerness} and purity-estimation~\cite{leone2023phase} protocols, they are useful to characterize quantum chaos in many-body states~\cite{leone2022stabilizer,oliviero2022black}, and are naturally related to important features of the so-called entanglement spectrum~\cite{tirrito2023quantifying,turkeshi2023measuring}. Therefore, we expect that the numerical techniques introduced in our work will be practically useful to compute the SRE in different contexts.

The rest of this manuscript is organized as follows. In Sec.~\ref{sec:stab_entropy} we introduce the stabilizer protocols, the SEs and the other monotones of interest in this work. In Sec.~\ref{sec:counter_example} we present our counterexamples to (strong) monotonicity for the SEs. In Sec.~\ref{sec:relations} we study order relations between the SEs and the other measures, both using analytical arguments and providing numerical evidence for small numbers of qubits. Finally, in Sec.~\ref{sec:numerics} we discuss efficient evaluation methods for SEs of MPSs, putting forward a novel approach to compute the von Neumann SE.  Our conclusions are consigned to Sec.~\ref{sec:conclusions}.

\section{Stabilizer Protocols and monotonicity}
\label{sec:stab_entropy}

\subsection{Stabilizer protocols}

We consider a system of $N$ qubits, with Hilbert space $\mathcal{H}=\otimes_{j=1}^N\mathcal{H}_j$, and $\mathcal{H}_j\simeq \mathbb{C}^{2}$. We denote by $\{\sigma^\alpha\}_{\alpha=0}^3$ the Pauli matrices ($\sigma^{0}=\openone$) and by $\{\ket{0}$, $\ket{1}\}$ the local computational basis. Denoting by $\tilde{\mathcal{P}}_N=\{i^{\alpha_0}\bigotimes_{k=1}^N\sigma^{\alpha_k}\}_{\alpha_0,\alpha_1,\dots,\alpha_N}$ the Pauli group, i.e. the group of all $N$-qubit Pauli strings with phases $\pm 1$, $\pm i$, we define the Clifford group to be the set of unitary operators $U$ such that $U WU^\dagger\in \tilde{\mathcal{P}}_N$ for all $W\in \tilde{\mathcal{P}}_N$. Then, pure stabilizer states are the states generated by applying elements of the Clifford group to the reference state $\ket{0}^{\otimes N}$. Finally, we also introduce $\mathcal{P}_N=\{\bigotimes_{k=1}^N\sigma^{\alpha_k}\}_{\alpha_1,\dots,\alpha_N}$ as the set of Pauli strings with trivial phase $+1$ only.

According to standard resource theory~\cite{chitambar2019quantum}, in order to define nonstabilizerness (or magic) monotones, one needs to first specify the set of free operations. This choice is not unique~\cite{liu2022many,heimendahl2022axiomatic}, but a minimal set, typically included in the set of free operations, is that of  \emph{stabilizer protocols}. These are quantum channels consisting of the following elementary operations:
\begin{enumerate}
	\item Clifford unitary operations, $\rho\mapsto U \rho U^\dagger$, where $U$ is in the Clifford group;
	\item Composition with stabilizer states, $\rho\mapsto \rho \otimes S$, where $S$ is a stabilizer state;
	\item Measurements in the computational basis; \label{item:measurements}
	\item Discarding of qudits; \label{item:partial_trace}
	\item The above operations conditioned on the outcomes of measurements.
\end{enumerate}
Given a stabilizer protocol $\mathcal{E}$, a nonstabilizerness monotone $M$ then satisfies
\begin{equation}
	\label{eq:monotonicity_condition}
M\left[\mathcal{E}(\rho)\right]\leq M(\rho)\,.
\end{equation}
In addition to the above, one can define a further property which is sometimes referred to as strong monotonicity~\cite{chitambar2019quantum}. Consider performing computational-basis measurements on a subset $\Lambda$ containing $m$ qubits. We denote by ${\lambda}=\{\lambda_1,\ldots, \lambda_m\}$ the set of outcomes for a single measurement ($\lambda_j=0,1$), by $\rho_{\bf \lambda}$ the post-measurement state, and by $p_{\lambda}={\rm Tr}\left[ \Pi_\lambda \rho \Pi_\lambda \right]$ the corresponding probability, where 
\begin{equation}
\Pi_\lambda= |\lambda\rangle \langle \lambda| \otimes \openone_{N\setminus m}\,.
\end{equation}
The strong-monotonicity condition, with respect to computational-basis measurements, read
\begin{equation}
	M(\rho)\geq \sum_{\lambda}p_\lambda M[\rho_\lambda]\,.
	\label{eq:strong_monotonicity}
\end{equation} 
This relation states that on average the nonstabilizerness cannot increase due to computational-basis measurements. Eq.~\eqref{eq:strong_monotonicity} is not required for a function to be a magic monotone. For instance, the min-relative entropy (introduced in the next section) does not satisfy it. Still, it seems particularly desirable in the many-body setting, as we will discuss later on. 

As mentioned, this work is restricted to pure states. Therefore, given a quantum state $\ket{\psi}$, in order to test monotonicity of a measure $M$, we need to consider stabilizer protocols such that
\begin{equation}
	\mathcal{E}(\ket{\psi}\bra{\psi})=\ket{\phi}\bra{\phi}\,,
\end{equation}
for some state $\ket{\phi}$. This corresponds to a deterministic state-transformation protocol. Crucially,  $\mathcal{E}$ is not necessarily a unitary channel, and could feature measurements and subsequent Clifford operations conditioned on the outcomes. These feedback operations are necessary in order to make the transformation deterministic, given the randomness of the measurement outcomes. We will discuss an explicit example in Sec.~\ref{sec:violation_mono}.

\subsection{SEs and monotones}

We now introduce the SEs and two known nonstabilizerness monotones, the min-relative entropy and the robustness of magic. Given a pure state $\ket{\psi}\in\mathcal{H}$, the R\'enyi SE of order $n$ reads~\cite{leone2022stabilizer}
\begin{equation}\label{eq:SRE}
	M_{n}(|\psi\rangle)=(1-n)^{-1} \log \sum_{P \in \mathcal{P}_{N}} \frac{\braket{\psi|P|\psi}^{2n}}{2^N}\,,
\end{equation}
where we use the natural logarithm. We note that it can be rewritten as~\cite{leone2022stabilizer}
\begin{equation}
M_{n}(|\psi\rangle)=(1-n)^{-1} \log \sum_{P \in \mathcal{P}_{N}} \Xi_{P}^{n}(|\psi\rangle)-N\log 2\,,
\end{equation}
where $\Xi_{P}(|\psi\rangle)=\braket{\psi|P|\psi}^{2}/2^{N}$. Since $\Xi_{P}(|\psi\rangle)\geq 0$ and $\sum_{P \in \mathcal{P}_{N}}\Xi_{P}(|\psi\rangle)=1$, it can be interpreted as the R\'enyi-$n$ entropy of the classical probability distribution $\Xi_{P}(|\psi\rangle)$, up to an off-set  $-N\log 2$. The von Neumann SE is obtained by taking the limit $n\to1$.

Next, we introduce the so-called log-free robustness of magic~\cite{howard2017application,liu2022many}, which is defined as~\cite{howard2017application}
\begin{equation}
	\text{LR}(\rho)=\log\left[\text{min}_x\left\{ \sum_i \vert x_i \vert: \rho=\sum_i x_i \sigma_i\right\}\right]\,,
\end{equation}
where $S=\{\sigma_i \}$ is the set of pure $N$-qubit stabilizer states. Note that $\rho$ can be either a pure or a mixed state. The robustness of magic is then defined as
\begin{equation}
R(\rho)=e^{\text{LR}(\rho)}\,.
\end{equation}

Finally, we consider the min-relative entropy of magic~\cite{bravyi2019simulation,liu2022many}. For a given pure state $\ket{\psi}$, it reads
\begin{equation}
	D_\text{min}(\ket{\psi})=-\log\left[F_\text{STAB}(\ket{\psi})\right]\,,
\end{equation}
where we introduced the stabilizer fidelity
\begin{equation}
F_\text{STAB}(\ket{\psi})=\text{max}_{\ket{\phi}}\left\{\vert\braket{\psi\vert\phi}\vert^2\right\}\,,
\end{equation}
where the maximum is taken over the set of stabilizer states $\ket{\phi}$. Roughly speaking, $D_\text{min}$ measures the distance between $\ket{\psi}$ and its nearest stabilizer state. 

The robustness and the min-relative entropy of magic are known to be genuine nonstabilizerness monotones~\cite{howard2017application,liu2022many}. In addition, the robustness of magic satisfies the strong monotonicity condition~\cite{regula2017convex}, while the min-relative entropy does not. Both monotones are natural objects from the resource-theory point of view, and it was argued in Ref.~\cite{liu2022many} that any ``good'' magic measure $M$ should satisfy
\begin{equation}
 D_{\rm min}(\rho)\leq M(\rho)\leq {\rm LR}(\rho)\,.
\end{equation}

So far, the question of whether the SEs are monotones with respect to stabilizer protocols remained open. In the next section, we will provide an explicit counterexample showing that, for R\'enyi index $0\leq n< 2$, they are not. In addition, we will show that, for any R\'enyi index, the SEs are not strong monotones. Despite these results, the SEs could still be of interest from the resource-theory point of view. For instance, we will discuss in Sec.~\ref{sec:relations} order relations between the SEs and known monotones. Given the possibility to efficiently compute the SEs, cf. Sec.~\ref{sec:numerics}, this could allow one to obtain useful bounds on the nonstabilizerness resource of a given state. 

\section{Counterexamples to monotonicity}
\label{sec:counter_example}

\subsection{Violation of monotonicity for R\'enyi index $0\leq n < 2$}
\label{sec:violation_mono}

We start by presenting our counterexample to the monotonicity condition~\eqref{eq:monotonicity_condition} of the SEs for R\'enyi index $0\leq n<2$. In Appendix~\ref{sec:details_counter}, we will discuss in detail how this counterexample was found. 

Consider a set of $N=4$ qubits, and take the ordered computational basis 
\begin{align}
\mathcal{B}=\{\ket{0000}, \ket{0001}, \ket{0010}, \ket{0011}, \ket{0100}\,,\nonumber\\
 \ldots ,\ket{1110},\ket{1111}\}\,.
\end{align}
We will consider the qubits to be ordered from left to right, so that qubit $1$ is the leftmost, etc. We define the state $\ket{\varphi^\ast}$ by its coordinates in the basis $\mathcal{B}$, which read
\begin{align}
\frac{1}{2\sqrt{6}}\{&0, 0,  0, 2i, 1-i,-1-i,-1+i, -1+i,\nonumber\\
	&\sqrt{2}, \sqrt{2}i, -\sqrt{2}, -\sqrt{2}, 0, 0, 0,\sqrt{2}(1+i)
	\}\,,\label{eq:counter_example}
\end{align}
so that we have, for instance,
\begin{align}
	\braket{0000|\varphi^\ast}&=\braket{1110|\varphi^\ast}=0,\\
	\braket{0011|\varphi^\ast}&=i/\sqrt{6}\,,
\end{align}
etc. Next, consider the protocol defined by the following set of instructions: 
\begin{enumerate}
\item Perform a projective measurement (in the computational basis) of the first (leftmost) qubit;
\item If the outcome is $1$, do nothing;
\item If the outcome is $0$, apply the following unitary operator
\begin{equation}
	\label{eq:unitary}
	U=V_2 V_1\,,
\end{equation}
where
\begin{align}
	V_1&=\sigma^x_1\otimes \frac{(\sigma^x_2-\sigma^y_2)}{\sqrt{2}}\otimes \openone_3\otimes \openone_4\,,\\
	V_2&=\openone_1\otimes \openone_2\otimes  w_{34} \,,
\end{align}
and
\begin{equation}
	w_{34}=(\sigma^x_3\otimes \openone_4) {\rm CZ}_{34} (\sigma^x_3\otimes \openone_4)\,.
\end{equation}
\end{enumerate}
Here ${\rm CZ}_{34}$ is the (Clifford) controlled-Z gate. Namely, in the basis $\{\ket{00}_{34}, \ket{01}_{34}, \ket{10}_{34}, \ket{11}_{34}\}$, we have ${\rm CZ}_{34}={\rm diag}(1,1,1,-1)$.  

Note that this is a stabilizer protocol, because all the unitary operators applied after the measurement are Cliffords. In particular, the operator $(\sigma^x-\sigma^y)/\sqrt{2}$ is a Clifford operator, as it can be seen from the decomposition into Hadamard ($H$) and $S$ gates,
\begin{equation}
(\sigma^x-\sigma^y)/\sqrt{2}=HSHSH\sigma^x\,,
\end{equation} 
where $S={\rm diag}(1,i)$.

Let us apply the protocol $\mathcal{E}$ to the input state $\ket{\varphi^\ast}$. A priori, $\mathcal{E}(\ket{\varphi^\ast}\bra{\varphi^\ast})$ is a mixed state, because of the random outcome of the measurement. However, the feedback unitary operator~\eqref{eq:unitary} has been chosen to make the output state pure (this is true only when the input state is $\ket{\varphi^\ast}$), i.e. the output state is independent of the measurement outcome and the state-transformation protocol is deterministic. In particular, by an explicit calculation, one can show $\mathcal{E}(\ket{\varphi^\ast}\bra{\varphi^\ast})=\ket{\psi^\ast}\bra{\psi^\ast}$ where
\begin{equation}\label{eq:product_final}
\ket{\psi^\ast}=\ket{1}_1\otimes \ket{\chi^\ast}_{234}\,,
\end{equation}
and
\begin{align}
\ket{\chi^\ast}=\frac{1}{\sqrt{6}}\left(\ket{000}+i\ket{001}-\ket{010}-\ket{011}\right.\nonumber\\
\left. +(1+i)\ket{111}
\right)\,.
\end{align}
Note in particular that
\begin{equation}\label{eq:psi_star_pi1}
	\ket{\psi^\ast}=\frac{1}{\sqrt{p_1}}\Pi_1\ket{\varphi^\ast}\,,
\end{equation}
where $p_1=\braket{\varphi^\ast|\Pi_1|\varphi^\ast}$, and $\Pi_1=\ket{1}\bra{1}\otimes \openone_{234}$. 

Finally, we define the difference between the SEs of the input and output state:
\begin{align}
\Delta M(n):&=M_n(\ket{\varphi^\ast})-M_n(\ket{\psi^\ast})\nonumber\\
&=M_n(\ket{\varphi^\ast})-M_n(\ket{\chi^\ast})\,.
\label{eq:delta_m}
\end{align}

\begin{figure}
	\includegraphics[scale=0.52]{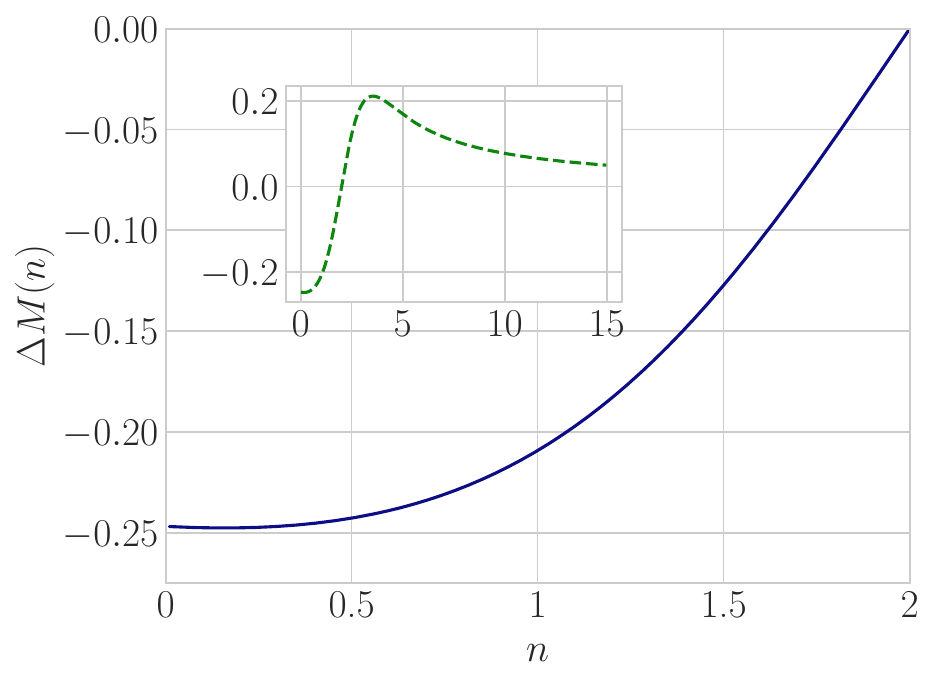}
	\caption{Numerical evaluation of the difference $\Delta M(n)$ defined in Eq.~\eqref{eq:delta_m}, showing $\Delta M(n)<0$ for $0\leq n <2$. Inset: same plot for $n\in [0,15]$.}
	\label{fig:deltam}
\end{figure}

In the second line, we have used the factorization~\eqref{eq:product_final}, the fact that the SE is additive and that $M_n(\ket{1})=0$. The function $\Delta M(n)$ can now be straightforwardly evaluated numerically. We plot it in Fig.~\ref{fig:deltam} for $n\in[0,15]$, from which we see that $\Delta M(n)<0$ for $0\leq n<2$, implying a violation of the monotonicity condition~\eqref{eq:monotonicity_condition}. 

A few comments are in order. First, note that, while not being a stabilizer state, $\ket{\varphi^\ast}$ is an eigenstate of the Clifford operation $U$ in~\eqref{eq:unitary}, i.e.
\begin{equation}
	\ket{\varphi^\ast}=U\ket{\varphi^\ast}\,.
\end{equation}
Ultimately, this makes it possible to find a Clifford feedback operation which makes the state-transformation deterministic, because
\begin{align}
U\ket{0}\bra{0}_1\ket{\varphi^\ast}=&U\frac{(1+\sigma^z_1)}{2}\ket{\varphi^\ast}\nonumber\\
=&\frac{(1-\sigma^z_1)}{2}U\ket{\varphi^\ast}=\ket{1}\bra{1}\ket{\varphi^\ast}\,,
\end{align}
where we used $U\sigma^z_1=-\sigma^z_1U$\,.

For $n\geq 2$, we were not able to find examples with $\Delta M(n)<0$ for $N=4$ qubits. As we explain in Appendix~\ref{sec:details_counter}, the state $\ket{\varphi^\ast}$ was found by maximizing numerically the violation of the strong monotonicity condition~\eqref{eq:strong_monotonicity}. For $N=4$, we find violations for $0\leq n <2$. Increasing $N$, we find violations even for $n\geq 2$, cf. also Sec.~\ref{sec:violation}. For instance, for $N=5$ we find that the strong monotonicity condition can be violated up to $n\simeq 3.2$. The latter, however, does not imply violation of~\eqref{eq:monotonicity_condition}. In particular, none of the non-trivial states that we have found for $N=5$ violating the strong monotonicity condition are eigenstates of Clifford operators, so that we could not devise a deterministic protocol as the one presented in this section. Therefore, the question of whether the SEs are monotones for R\'enyi index $n\geq 2$ (at least when restricted to pure states) remains open.

\subsection{Violation of strong monotonicity}
\label{sec:violation}

We now show that, for any value of $n$, the SEs do not satisfy the strong monotonicity condition~\eqref{eq:strong_monotonicity}.  First, we show this for $0\leq n<2$. To this end, it is enough to consider the counterexample~\eqref{eq:counter_example}. Indeed, let us define
\begin{equation}
	\ket{\varphi^\ast_0}=\frac{1}{\sqrt{p_0}}\Pi_0\ket{\varphi^\ast}\,,\quad 	\ket{\varphi^\ast_1}=\frac{1}{\sqrt{p_1}}\Pi_1\ket{\varphi^\ast}\,,
\end{equation}
where $\Pi_{a}=\ket{a}\bra{a}\otimes \openone_{234}$ and $p_a=\braket{\varphi^\ast|\Pi_a|\varphi^\ast}$. It is straightforward to show
\begin{equation}
	M_{n}(\ket{\varphi^\ast_0})=M_{n}(\ket{\varphi^\ast_1})\,.
\end{equation}
Therefore, because of~\eqref{eq:psi_star_pi1}, we have
\begin{equation}
	p_0M_n(\ket{\varphi^\ast_0})+p_1M_n(\ket{\varphi^\ast_1})=M_n(\ket{\psi^\ast})\,.
\end{equation}
Using the results of Sec.~\ref{sec:violation_mono}, we have $M_n(\ket{\psi^\ast})>M_n(\ket{\varphi^\ast})$ for $0\leq n<2$. This proves that strong monotonicity is violated for $0\leq n<2$.

Next, let us show that the condition~\eqref{eq:strong_monotonicity} is also violated for $n\geq 2$. To this end, let us consider a system of $N$ qubits, and define the state
\begin{equation}\label{eq:psi_eps_state}
\ket{\psi^\varepsilon}=\frac{1}{\sqrt{\mathcal{N}_\varepsilon}}\left[\ket{0}^{\otimes N}+ \varepsilon\ket{\chi}^{\otimes N}\right]\,,
\end{equation}
where $\ket{\chi}$ is the magic state~\cite{bravyi2005universal}
\begin{align}\label{eq:t_state}
	\ket{\chi}&=e^{-i(\pi / 4)}\cos \beta|0\rangle+ \sin \beta|1\rangle\,,
\end{align}
with $\cos(2\beta)=1/\sqrt{3}$, and 
\begin{equation}\label{eq:norm}
	\mathcal{N}_\varepsilon=1+\varepsilon^2+2\varepsilon \cos( \beta)^N[\cos(N \pi / 4)]\,.
\end{equation}
Note that
\begin{equation}
	\ket{\chi}\bra{\chi}=\frac{1}{2}\left[\openone_1+\frac{1}{\sqrt{3}}\left(\sigma^x+\sigma^y+\sigma^z\right)\right]\,.
\end{equation}
We claim that, for finite $\varepsilon>0$ and $n>1$, we have
\begin{equation}\label{eq:const_bound}
M_n(\ket{\psi^\varepsilon})<c_n(\varepsilon)\,, \qquad (n>1)
\end{equation}
where $c_n(\varepsilon)$ is a constant which depends on $n$ and $\varepsilon$, but not on $N$. We prove this in Appendix~\ref{sec:details_counter}, where we also support this statement numerically. 

Consider now performing a projective measurement of the first qubit. As usual, we denote by $\ket{\psi^\varepsilon_0}$, $\ket{\psi^\varepsilon_1}$ the post-measurement states associated with the outcomes $0$ and $1$, and by $p_0$, $p_1$ the corresponding probabilities. When the measurement outcome is $1$, the post-measurement state is 
\begin{equation}
\ket{\psi^\varepsilon_1}=\ket{1}_1\otimes\ket{\chi}^{\otimes (N-1)}\,.
\end{equation}
Because of additivity and the fact that $M_n(\ket{1})=0$, we have
\begin{equation}
M_n(\ket{\psi^\varepsilon_1})=(N-1)M_n(\ket{\chi})\,,
\end{equation}
so that
\begin{align}
	p_0M_n(\ket{\psi^\varepsilon_0})+p_1M_n(\ket{\psi^\varepsilon_1})&\geq p_1M_n(\ket{\psi^\varepsilon_1})\nonumber\\
	&= p_1 (N-1) M_n(\ket{\chi})\,.\label{eq:p_j}
\end{align}
Finally, we note that $p_1$ remains finite in the thermodynamic limit $N\to \infty$, 
\begin{equation}\label{eq:limit}
	\lim_{N\to\infty}	p_1=\lim_{N\to\infty}\frac{1}{\mathcal{N}_\varepsilon}\varepsilon^2\sin\beta^2=\frac{\varepsilon^2}{1+\varepsilon^2}\sin\beta^2\,.
\end{equation}
Combining Eqs.~\eqref{eq:const_bound},~\eqref{eq:p_j}, and~\eqref{eq:limit}, we obtain a violation of the strong monotonicity condition for sufficiently large $N$ and $n>1$ (and thus also for $n\geq 2$). 

In passing, we mention that this example can also be used to show that the min-relative entropy $D_\text{min}$ is not a strong monotone. Moreover, it illustrates how, in the many-body setting, strong monotonicity appears to be particularly important. Indeed, if strong monotonicity is absent, we see here that  measuring a single qubit is enough to increase nonstablizerness from a $O(1)$ constant to a $O(N)$ value with finite probability.

Finally, let us mention that the results presented in this section contradict previous claims in the literature. This is because the R\'enyi-$1/2$ SE coincides with the measure introduced in Ref.~\cite{hahn2022quantifying}, which was claimed to satisfy the strong monotonicity condition~\eqref{eq:strong_monotonicity}, cf. also~\cite{hahn2023erratum}. 

\section{Relations to other monotones}
\label{sec:relations}

As mentioned, even if they fail to be monotones,  SEs could still be of interest from the resource-theory point of view, based on their relation with known monotones. In this section we discuss elementary inequalities between the SEs, the min-relative entropy and the robustness of magic (as usual, restricting to pure states). Note that the latter two are known to satisfy the general inequality~\cite{liu2022many}
\begin{equation}\label{eq:d_LR}
	D_{\rm min}(\ket{\psi})\leq {\rm LR}(\ket{\psi})\,.
\end{equation}
First, we recall the following inequality~\cite{howard2017application,leone2022stabilizer}
\begin{equation}\label{eq:mn_lr}
	M_{n}(\ket{\psi})\leq 2 {\rm LR}(\ket{\psi})\, \qquad (n\geq 1/2)\,.
\end{equation}
For $n >1$ we now derive an upper bound also in terms of the min-relative entropy, namely
\begin{equation}\label{eq:to_prove}
M_n(\ket{\psi})\leq \frac{2n}{n-1}D_{\rm min}(\ket{\psi}) \,.
\end{equation}
To see this, we note that for any state $\ket{\psi}$ with stabilizer fidelity $F_\text{STAB}(\ket{\psi})$, we can find a Clifford unitary $U_C$ s.t.
\begin{equation}\label{eq:dec}
\ket{\phi}=U_C\ket{\psi}=\sum_k a_k \ket{k}
\end{equation}
 where 
\begin{equation}\label{eq:boundFmax}
	\vert a_0\vert^2 = F_\text{STAB}(\ket{\psi})\,.
\end{equation}
This follows immediately from $F_\text{STAB}=\vert \braket{\psi\vert\phi_\text{max}}\vert^2=\vert \braket{\psi\vert U_C^\text{max}\vert 0}\vert^2=\vert a_0\vert^2$. In Eq.~\eqref{eq:dec}, we have denoted by $\{\ket{n}\}$ the set of states in the computational basis. Next, we have
\begin{align}\label{eq:inequality}
	&2^{-N}\sum_{P \in \mathcal{P}_N} \vert \bra{\phi}P\ket{\phi}\vert^{2n}\ge 2^{-N}
	\sum_{P \in \mathcal{P}_z} \vert \bra{\phi}P\ket{\phi}\vert^{2n}\nonumber\\
	&\ge
	2^{-2n N} \left(\sum_{P \in \mathcal{P}_z} | \bra{\phi}P\ket{\phi}|\right)^{2n}\,.
\end{align}
Here $\mathcal{P}$ is the set of Pauli strings while $\mathcal{P}_z$ is the set of Pauli strings containing $\openone$ and $\sigma^z$ only.
In the second line we used the convexity inequality
\begin{align}\label{eq:powermean}
	\sum_{i=1}^m \vert a_i\vert^k\ge \frac{1}{m^{k-1}}\left(\sum_{i=1}^m \vert a_i \vert \right)^k\,.
\end{align}
Now, using
\begin{equation}
	\left(\sum_{P \in \mathcal{P}_z} | \bra{\phi}P\ket{\phi}|\right)^{2n}\geq \left|  \sum_{P \in \mathcal{P}_z} \bra{\phi}P\ket{\phi}\right|^{2n}
\end{equation}
and
\begin{equation}
	\sum_{P\in \mathcal{P}_z}P=2^{N}\ket{0}\bra{0}\,,
\end{equation}
Eq.~\eqref{eq:inequality} yields
\begin{equation}\label{eq:almost_there}
2^{-N}\sum_{P \in \mathcal{P}_N} \vert \bra{\phi}P\ket{\phi}\vert^{2n}\geq |a_0|^{4n}=F_{\rm STAB}(\ket{\psi})^{2n}\,.
\end{equation}
Finally, using that the SE is invariant under unitary Clifford operations, we have $M_n(\ket{\psi})=M_n(\ket{\phi})$ and, combining with Eq.~\eqref{eq:almost_there}, we arrive at Eq.~\eqref{eq:to_prove}.

Given the inequalities~\eqref{eq:d_LR} and~\eqref{eq:to_prove}, it would be very useful to also provide an $N$-independent lower bound of the SE in terms of either the log-robustness or the min-relative entropy. In the following, we provide a simple example showing that this is not possible for the log-robustness of magic if $n>1/2$. Consider the single-qubit state
\begin{equation}
	\ket{\Omega(s)}=\cos(s)\ket{0}+\sin(s)\ket{1}\,.
\end{equation} 
For small $s$, we can compute
\begin{align}
	M_{n}(\ket{\Omega(s)})&=a_n s^2+o(s^2)\,,\quad n>1\,,\\
	M_{n}(\ket{\Omega(s)})&=b_n s^{2n}+o(s^{2n})\,,\quad n<1\,,
\end{align}
where $a_n$, $b_n$ are $n$-dependent constants. Now, introducing the $N$-qubit state
\begin{equation}
	\ket{\Lambda_N(s_0)}=\ket{\Omega(s_0/\sqrt{N})}^{\otimes N}\,,
\end{equation}
and using additivity of the SE, we have
\begin{align}
	M_{n}(\ket{\Lambda_N(s_0)})&=a_n s_0^2+o(1)\,,\quad n>1\,,\\
	M_{n}(\ket{\Lambda_N(s_0)})&=b_{n} s^{2n}_0 N^{1-n}+o(N^{1-n}) ,\quad n<1\,.
\end{align}
Therefore, using~\eqref{eq:mn_lr}, we have that, for $n>1/2$
\begin{equation}
\frac{	{\rm LR}(\ket{\Lambda_N(s_0)})}{M_{n}(\ket{\Lambda_N(s_0)})}\geq \frac{1}{2}\frac{M_{1/2}(\ket{\Lambda_N(s_0)}}{M_{n}(\ket{\Lambda_N(s_0)}}\sim N^{\text{min}(n-\frac{1}{2},\frac{1}{2})}\,.
\end{equation}
This shows that it is not possible to derive an $N$-independent lower bound for $n>1/2$. 

On the other hand, since $	D_{\rm min}(\ket{\Omega(s)})\propto s^2$, and given the sub-additivity of the min-relative entropy, this example does not rule out the possibility that the SE with $n>1$ can be lower-bounded by $D_{\rm min}$. Since $M_n$ vanishes for $n\to\infty$, this bound should necessarily involve a $n$-dependent factor. In order to avoid dealing with this, we have performed extensive numerical minimization of the fraction $M_n(\ket{\psi})/D_\text{min}(\ket{\psi})$ for small values of $n$ up to $N=4$ qubits. We have found numerical evidence that the minimized value of the ratio appears to be a constant, which only depends on $n$. In particular, we found $M_n(\ket{\psi})\gtrsim 1.7D_\text{min}(\ket{\psi})$ for all states $\ket{\psi}$ of up to $N=4$ qubits and $1\le n \le2$. For higher $n$, the minimized value for $N\le4$ appears to scale as $O(1/n)$, which matches the overall scaling of $M_n\sim O(1/n)$ for $n\gg1$. Overall, our results hint at the possibility that such a lower bound exists. We leave this as an open question for future research. 

\section{Numerical methods for the SE}
\label{sec:numerics}

One of the reasons why SEs are appealing is the fact that they can often be computed much more efficiently than previously known nonstabilizerness monotones.  In fact, even if for generic states the computational cost grows exponentially in $N$~\cite{leone2022stabilizer,oliviero2022magic}, the evaluation of Pauli expectation values is, in practice, simpler than carrying out the minimization procedure involved in the definition of the robustness and min-relative entropy of magic. In addition, for certain classes of many-body states, the computational cost is polynomial in the number of qubits $N$. This is the case for MPSs~\cite{perez2007matrix,cirac2017matrix_op,cirac2020matrix}, the simplest example of tensor-network states~\cite{silvi2019tensor}. This was pointed out in Ref.~\cite{haug2022quantifying}, where an efficient computational method was put forward for the computation of R\'enyi SEs with integer index $n>1$.  

The aim of this section is to review the result of Ref.~\cite{haug2022quantifying}, and complement it putting forward an approach to the computation of the von Neumann SE. Combined with the bounds discussed in Sec.~\ref{sec:relations},  these methods could be practically useful from the resource-theory point of view, when the number of qubits makes the computation of the min-relative entropy or robustness of magic unfeasible. 

First, we recall the definition of an MPS (with open boundary conditions) $\ket{\Psi_N}$~\cite{perez2007matrix}
\begin{equation}\label{eq:MPS}
	\ket{\Psi_{N}}=\sum_{\{s_{k}\}} A^{s_{1}}_1 \ldots A^{s_{N}}_N\ket{s_{1}, \ldots, s_{N}}\,,
\end{equation}
where $A^{s}_{k}$ are $\chi_{k}\times \chi_{k+1}$ matrices, with $\chi_1=\chi_{N+1}=1$. We call $\chi=\max_{j}\chi_j$ the \emph{bond-dimension}.
The method developed in Ref.~\cite{haug2022quantifying} starts from the simple identity
\begin{align}\label{eq:contraction}
	\sum_{P \in \mathcal{P}_{N}} \frac{\braket{\Psi_N|P|\Psi_N}^{2n}}{2^N}&=
	(\bra{\Psi_N}\otimes \bra{\overline{\Psi}_N})^{\otimes n}\Lambda^{(n)}_1\otimes \Lambda^{(n)}_2\otimes\nonumber\\
	&\cdots \otimes \Lambda^{(n)}_N (\ket{\Psi_N}\otimes \ket{\overline{\Psi}_N})^{\otimes n}\,,
\end{align}
where $\Lambda^{(n)}_j=(1/2)\sum_{\alpha=0}^3 (\sigma^\alpha_j\otimes \overline{\sigma^{\alpha}}_j )^{\otimes n}$, while $\overline{(\cdot)}$ denotes complex conjugation. This formula allows to one replace the sum over exponentially many terms with the computation of the norm of an MPS of bond-dimension $\chi^{2n}$, which can be carried out efficiently in $N$, at a cost $O(N\chi^{6n})$~\cite{haug2022quantifying}. This mapping is also convenient from the analytic point of view. For instance, it allows one to show that the density of SE can be extracted locally for MPSs without long-range correlations. 

This method does not allow for a numerical evaluation of the SE with arbitrary R\'enyi index $n$. However, for $n=1$ an efficient numerical scheme is possible exploiting once again the properties of MPSs. To this end, we start from the expression
\begin{equation}\label{eq:m1}
	M_1(\ket{\psi})=-\sum_{P\in\mathcal{P}_N}\Xi_P(\ket{\psi})\log \Xi_P(\ket{\psi})-N \log 2\,,
\end{equation}
and interpret $\Xi_P(\ket{\psi})$ as a probability distribution. Since each term in the sum is positive, we can evaluate $M_1(\ket{\psi})$ by sampling the probability distribution $\Xi_P(\ket{\psi})$. Crucially, for an MPS $\ket{\Psi_N}$, we do not need a Monte Carlo approach, but we can sample from $\Xi_P(\ket{\Psi_N})$ exactly, slightly generalizing the algorithm of perfect MPS sampling of Ref.~\cite{ferris2012perfect}. 

\begin{figure*}[htbp]
	\centering	
 	\subfigimg[width=0.3\textwidth]{a}{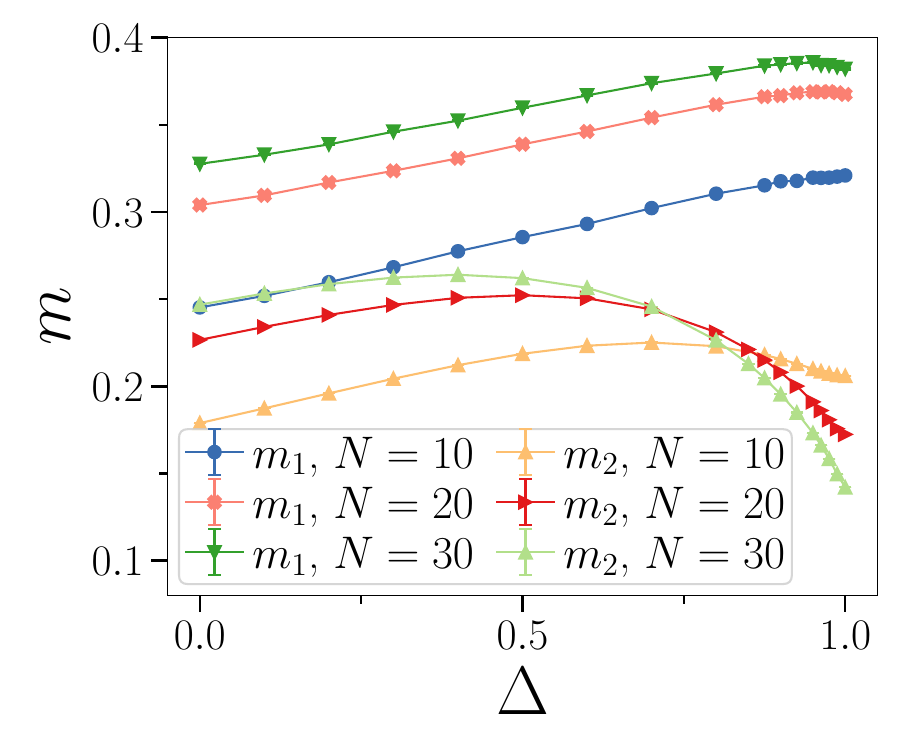}
	\subfigimg[width=0.3\textwidth]{b}{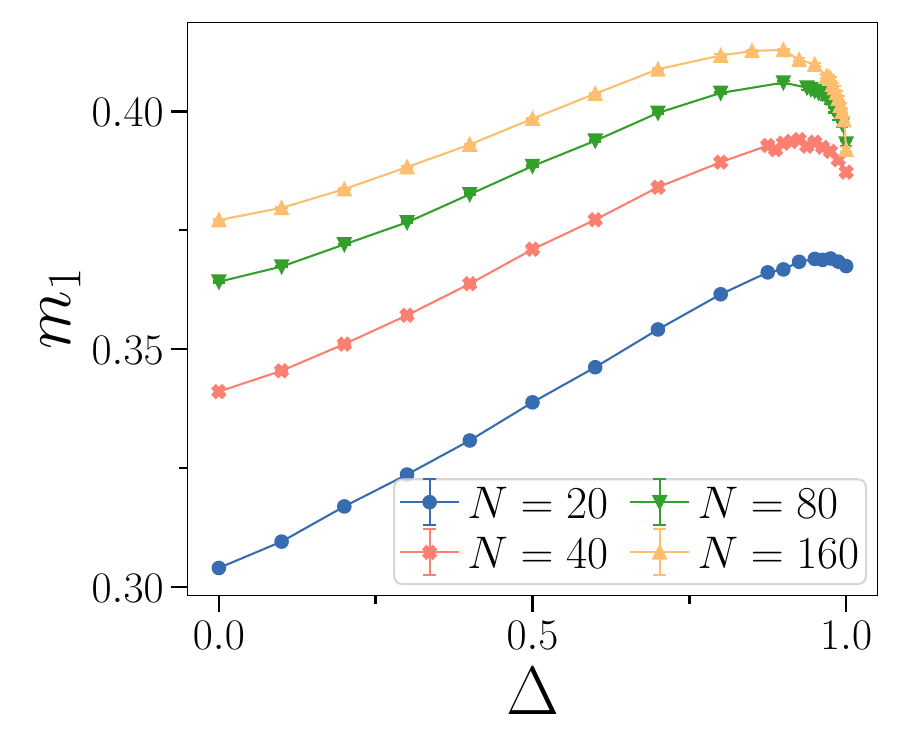}
	\caption{
  \idg{a} Von Neumann SE density $m_1$ and R\'enyi-$2$ SE density against anisotropy $\Delta$ for the ground state of the XXZ model, at half-filling. For $m_1$, the bond dimension is $\chi=40$, while the number of samples is $S=10^5$. For $m_2$, we used the method of Ref.~\cite{haug2022quantifying} with $\chi=12$.
 \idg{b} Von Neumann SE density $m_1$ against anisotropy $\Delta$ for the ground state of the XXZ model at half-filling. Here, $\chi=40$ and $S=10^5$.
	}
	\label{fig:magic}
\end{figure*}

Let us define $\boldsymbol{\alpha}=(\alpha_1,\ldots \alpha_N)$, with $\alpha_j=0,1,2,3$. The key observation is that the probability of the Pauli string associated with $\boldsymbol{\alpha}$, $P_{\boldsymbol{\alpha}}=\sigma_1^{\alpha_1}\cdots \sigma_N^{\alpha_N}$, can be written as
\begin{equation}\label{eq:prob_expression}
\Xi_{P_{\boldsymbol{\alpha}}}=p(\alpha_1)p(\alpha_2|\alpha_1)p(\alpha_3|\alpha_1,\alpha_2)\cdots\,,
\end{equation}
where $p(\alpha_j|\alpha_1,\alpha_2,\ldots)$ is the conditional probability that the Pauli matrix $j$ is $\sigma_j^{\alpha_j}$, if the Pauli matrices $k$ with $k<j$ are $\sigma_1^{\alpha_1}$, $\sigma^{\alpha_2}_2$, etc. Importantly, these probabilities can be efficiently computed for MPSs. To see this, we write explicitly
\begin{align}\label{eq:conditional_prob}
p(\alpha_j|\alpha_1,\ldots ,\alpha_{j-1})&={\rm Tr}\left[\rho_{1,\ldots j}\sigma_j^{\alpha_1}\cdots \sigma_j^{\alpha_j}\right.\nonumber\\
\times&\left.\rho_{1,\ldots j}\sigma_j^{\alpha_1}\cdots \sigma_j^{\alpha_j}\right]\,,
\end{align}
where $\rho_{1,\ldots j}$ is the reduced density matrix over the first $j$ qubits. This formula can be verified making repeated use of the identity $(1/2)\sum_{\alpha}\sigma^{\alpha}(\cdot)\sigma^{\alpha}=\openone {\rm Tr}(\cdot)$. For an MPS with bond dimension $\chi$, we can use the standard prescriptions~\cite{silvi2019tensor,evenbly2022practical} to evaluate the optimal contraction cost of the tensor network associated with~\eqref{eq:conditional_prob}, which yields $O(\chi^3 N)$. Therefore, the conditional probabilities can be computed efficiently in $N$. Note that the dependence with $\chi$ is favorable compared to that needed in the exact computation of the R\'enyi SEs for integer $n$~\cite{haug2022quantifying}.

Given~\eqref{eq:prob_expression}, we can use directly the algorithm detailed in Ref.~\cite{ferris2012perfect} to generate strings of Pauli operators associated with $\boldsymbol{\alpha}$, according to a probability which is exactly $\Xi_{P_{\boldsymbol{\alpha}}}$. Then, we can evaluate $M_{1}(\ket{\Psi_N})$ by averaging the value $\log\Xi_{P_{\boldsymbol{\alpha}}}(\ket{\Psi_N})$ over all the generated samples. 

The accuracy of the method increases with number of samples, which we denote by $S$ (in order not to distinguish it from the number of qubits $N$). Denoting by $\varepsilon_S$ the difference between the average value, obtained by sampling the distribution $\Xi_P$, and the actual value of $M_1$, standard arguments yield $\varepsilon_S=O(1/\sqrt{S})$. In turn, this implies that $M_1$ can be computed efficiently, i.e. at a computational cost scaling polynomially in $N$, for fixed accuracy.

We briefly illustrate this method for the ground-state of the interacting XXZ Heisenberg Hamiltonian
\begin{equation}
H_\text{XXZ}=-\sum_{k=1}^{N-1}(\sigma^x_k\sigma^x_{k+1}+\sigma^y_k\sigma^y_{k+1}+\Delta\sigma^z_k\sigma^z_{k+1})\,,
\end{equation}
where $\Delta$ is the anisotropy parameter, while we choose open boundary conditions. Note that the total magnetization $C=\sum_{k=1}^N\sigma^z_k$ is conserved, i.e. $[H_\text{XXZ},C]=0$.

In Fig.~\ref{fig:magic}, we report an example of our data for the SE of the ground state of the Heisenberg model at half-filling, i.e. the ground state within the subspace of states with $C\ket{\psi}=0$. 
In Fig.~\ref{fig:magic}a, we plot the SE density $m_n=M_n/N$ against $\Delta$ for the von Neumann SE density $m_1$ and the R\'enyi-$2$ SE density. Qualitatively, both $m_1$ and $m_2$ behave similar as a function of $\Delta$, where $m$ is first increasing with $\Delta$, then decreases. Further, both $m_1$ and $m_2$ increase in value with $N$, where we expect them to converge to a constant for large $N$ as the SE is extensive. Quantitatively, the decrease in $m$ towards $\Delta\rightarrow1$ is more pronounced for $m_2$. In Fig.~\ref{fig:magic}b, we show $m_1$ for larger values of $N$, illustrating the efficiency of the method. We see that $m_1$ decreases near $\Delta=1$ more sharply with higher $N$. 

In Appendix~\ref{sec:numerics_appendix} we provide further details on the numerical accuracy on the values $m_1$ and $m_2$, as a function of the bond dimension $\chi$ and number of samples $S$.

Finally, we note that one could in principle extend the perfect sampling method to arbitrary R\'enyi SEs. However, we expect that, in general, the number of samples required to obtained $M_n(\ket{\psi})$ up to fixed accuracy could scale exponentially in the number of qubits $N$.

\section{Outlook}
\label{sec:conclusions}

We have studied different aspects of the SEs, comparing them against the min-relative entropy and the robustness of magic. For R\'enyi index $0\leq n<2$, we have shown that the SEs are not monotones with respect to generic stabilizer protocols, not even when restricting to pure states. In addition, we have shown that, for any R\'enyi index, the SEs do not satisfy the strong monotonicity condition with respect to computational-basis measurements,  contradicting a previous claim in the literature~\cite{hahn2022quantifying}, cf. also~\cite{hahn2023erratum}. Next, we have discussed, both analytically and numerically, some inequalities between the SEs, the min-relative entropy and the robustness of magic. Finally, we have presented available tensor-network methods for the efficient computation of SEs in the context of MPSs, reviewing the results of Ref.~\cite{haug2022quantifying} and putting forward a new method for the von Neumann SE.

Our work raises several questions. First, it would be interesting to either generalize our counterexample to show rigorously that the monotonicity condition~\eqref{eq:monotonicity_condition} is violated also for R\'enyi index $n\geq 2$, or, conversely, to prove it. Second, as we discuss in Sec.~\ref{sec:relations}, we believe it would be important to understand whether one can derive a $N$-independent lower bound for the SE in terms of the min-relative entropy, at least for some R\'enyi index. Indeed, even if the SE is not a monotone, this result, together with those presented in Sec.~\ref{sec:relations}, could be used to provide upper and lower bounds to a genuine nonstabilizerness monotone, the min-relative entropy. 

Finally, our results show that the SEs are not monotones with respect to general stabilizer protocols (at least for R\'enyi index $0\leq n<2)$, and in any case not strong monotones. Therefore, our work highlights the need to find genuine nonstabilizerness monotones which are efficient to compute, at least in restricted classes of pure states such as MPSs. In this context, a very natural question is whether such a monotone can be defined in terms of the Pauli spectrum~\cite{beverland2020lower}, similarly to the SEs. We hope that our work will motivate further research in this direction.\\

\emph{Note added}: While finalizing this manuscript, the work~\cite{lami2023quantum} appeared on the arXiv, where a similar method to compute SEs in MPSs was put forward.\\

\begin{acknowledgements}
We thank Hyukjoon Kwon, Giacomo de Palma,  and especially Ludovico Lami for inspiring discussions.
\end{acknowledgements}
% ---------------------------------------------------------------
% ---------------------------------------------------------------
%\bigskip

\bibliography{bibliographyLong}

\begin{thebibliography}{53}
\providecommand{\natexlab}[1]{#1}
\providecommand{\url}[1]{\texttt{#1}}
\expandafter\ifx\csname urlstyle\endcsname\relax
  \providecommand{\doi}[1]{doi: #1}\else
  \providecommand{\doi}{doi: \begingroup \urlstyle{rm}\Url}\fi

\bibitem[Aaronson and Gottesman(2004)]{aaronson2004improved}
Scott Aaronson and Daniel Gottesman.
\newblock Improved simulation of stabilizer circuits.
\newblock \emph{Phys. Rev. A}, 70:\penalty0 052328, Nov 2004.
\newblock \doi{10.1103/PhysRevA.70.052328}.
\newblock URL \url{https://link.aps.org/doi/10.1103/PhysRevA.70.052328}.

\bibitem[Anwar et~al.(2014)Anwar, Brown, Campbell, and Browne]{anwar2014fast}
Hussain Anwar, Benjamin~J Brown, Earl~T Campbell, and Dan~E Browne.
\newblock Fast decoders for qudit topological codes.
\newblock \emph{New J. Phys.}, 16\penalty0 (6):\penalty0 063038, 2014.
\newblock \doi{10.1088/1367-2630/16/6/063038}.
\newblock URL
  \url{https://iopscience.iop.org/article/10.1088/1367-2630/16/6/063038/meta}.

\bibitem[Beverland et~al.(2020)Beverland, Campbell, Howard, and
  Kliuchnikov]{beverland2020lower}
Michael Beverland, Earl Campbell, Mark Howard, and Vadym Kliuchnikov.
\newblock Lower bounds on the non-clifford resources for quantum computations.
\newblock \emph{Quantum Science Tech.}, 5\penalty0 (3):\penalty0 035009, 2020.
\newblock \doi{10.1088/2058-9565/ab8963}.
\newblock URL \url{https://doi.org/10.1088/2058-9565/ab8963}.

\bibitem[Bravyi and Kitaev(2005)]{bravyi2005universal}
Sergey Bravyi and Alexei Kitaev.
\newblock Universal quantum computation with ideal clifford gates and noisy
  ancillas.
\newblock \emph{Phys. Rev. A}, 71:\penalty0 022316, Feb 2005.
\newblock \doi{10.1103/PhysRevA.71.022316}.
\newblock URL \url{https://link.aps.org/doi/10.1103/PhysRevA.71.022316}.

\bibitem[Bravyi et~al.(2019)Bravyi, Browne, Calpin, Campbell, Gosset, and
  Howard]{bravyi2019simulation}
Sergey Bravyi, Dan Browne, Padraic Calpin, Earl Campbell, David Gosset, and
  Mark Howard.
\newblock Simulation of quantum circuits by low-rank stabilizer decompositions.
\newblock \emph{Quantum}, 3:\penalty0 181, 2019.
\newblock \doi{10.22331/q-2019-09-02-181}.
\newblock URL \url{https://quantum-journal.org/papers/q-2019-09-02-181/}.

\bibitem[Bu and Koh(2019)]{bu2019efficient}
Kaifeng Bu and Dax~Enshan Koh.
\newblock Efficient classical simulation of clifford circuits with
  nonstabilizer input states.
\newblock \emph{Phys. Rev. Lett.}, 123:\penalty0 170502, Oct 2019.
\newblock \doi{10.1103/PhysRevLett.123.170502}.
\newblock URL \url{https://link.aps.org/doi/10.1103/PhysRevLett.123.170502}.

\bibitem[Bu et~al.(2022)Bu, Garcia, Jaffe, Koh, and Li]{bu2022complexity}
Kaifeng Bu, Roy~J Garcia, Arthur Jaffe, Dax~Enshan Koh, and Lu~Li.
\newblock Complexity of quantum circuits via sensitivity, magic, and coherence.
\newblock \emph{arXiv:2204.12051}, 2022.
\newblock URL \url{https://doi.org/10.48550/arXiv.2204.12051}.

\bibitem[Campbell(2011)]{campbell2011catalysis}
Earl~T. Campbell.
\newblock Catalysis and activation of magic states in fault-tolerant
  architectures.
\newblock \emph{Phys. Rev. A}, 83:\penalty0 032317, Mar 2011.
\newblock \doi{10.1103/PhysRevA.83.032317}.
\newblock URL \url{https://link.aps.org/doi/10.1103/PhysRevA.83.032317}.

\bibitem[Campbell(2014)]{campbell2014enhanced}
Earl~T. Campbell.
\newblock Enhanced fault-tolerant quantum computing in $d$-level systems.
\newblock \emph{Phys. Rev. Lett.}, 113:\penalty0 230501, Dec 2014.
\newblock \doi{10.1103/PhysRevLett.113.230501}.
\newblock URL \url{https://link.aps.org/doi/10.1103/PhysRevLett.113.230501}.

\bibitem[Campbell et~al.(2012)Campbell, Anwar, and Browne]{campbell2012magic}
Earl~T. Campbell, Hussain Anwar, and Dan~E. Browne.
\newblock Magic-state distillation in all prime dimensions using quantum
  reed-muller codes.
\newblock \emph{Phys. Rev. X}, 2:\penalty0 041021, Dec 2012.
\newblock \doi{10.1103/PhysRevX.2.041021}.
\newblock URL \url{https://link.aps.org/doi/10.1103/PhysRevX.2.041021}.

\bibitem[Campbell et~al.(2017)Campbell, Terhal, and Vuillot]{campbell2017roads}
Earl~T Campbell, Barbara~M Terhal, and Christophe Vuillot.
\newblock Roads towards fault-tolerant universal quantum computation.
\newblock \emph{Nature}, 549\penalty0 (7671):\penalty0 172--179, 2017.
\newblock \doi{10.1038/nature23460}.
\newblock URL \url{https://www.nature.com/articles/nature23460}.

\bibitem[Chen et~al.(2022)Chen, Garcia, Bu, and Jaffe]{chen2022magic}
Liyuan Chen, Roy~J Garcia, Kaifeng Bu, and Arthur Jaffe.
\newblock Magic of random matrix product states.
\newblock \emph{arXiv:2211.10350}, 2022.
\newblock URL \url{https://doi.org/10.48550/arXiv.2211.10350}.

\bibitem[Chitambar and Gour(2019)]{chitambar2019quantum}
Eric Chitambar and Gilad Gour.
\newblock Quantum resource theories.
\newblock \emph{Rev. Mod. Phys.}, 91:\penalty0 025001, Apr 2019.
\newblock \doi{10.1103/RevModPhys.91.025001}.
\newblock URL \url{https://link.aps.org/doi/10.1103/RevModPhys.91.025001}.

\bibitem[Cirac et~al.(2017)Cirac, Perez-Garcia, Schuch, and
  Verstraete]{cirac2017matrix_op}
J~Ignacio Cirac, David Perez-Garcia, Norbert Schuch, and Frank Verstraete.
\newblock Matrix product density operators: Renormalization fixed points and
  boundary theories.
\newblock \emph{Ann. Phys.}, 378:\penalty0 100--149, 2017.
\newblock \doi{10.1016/j.aop.2016.12.030}.
\newblock URL
  \url{http://www.sciencedirect.com/science/article/pii/S0003491616303013}.

\bibitem[Cirac et~al.(2021)Cirac, Perez-Garcia, Schuch, and
  Verstraete]{cirac2020matrix}
J~Ignacio Cirac, David Perez-Garcia, Norbert Schuch, and Frank Verstraete.
\newblock Matrix product states and projected entangled pair states: Concepts,
  symmetries, theorems.
\newblock \emph{Rev. Mod. Phys.}, 93\penalty0 (4):\penalty0 045003, 2021.
\newblock \doi{10.1103/RevModPhys.93.045003}.
\newblock URL
  \url{https://journals.aps.org/rmp/abstract/10.1103/RevModPhys.93.045003}.

\bibitem[Eastin and Knill(2009)]{eastin2009restriction}
Bryan Eastin and Emanuel Knill.
\newblock Restrictions on transversal encoded quantum gate sets.
\newblock \emph{Phys. Rev. Lett.}, 102:\penalty0 110502, Mar 2009.
\newblock \doi{10.1103/PhysRevLett.102.110502}.
\newblock URL \url{https://link.aps.org/doi/10.1103/PhysRevLett.102.110502}.

\bibitem[Evenbly(2022)]{evenbly2022practical}
Glen Evenbly.
\newblock A practical guide to the numerical implementation of tensor networks
  i: Contractions, decompositions, and gauge freedom.
\newblock \emph{Frontiers in Applied Mathematics and Statistics}, 8:\penalty0
  806549, 2022.
\newblock \doi{10.3389/fams.2022.806549}.
\newblock URL \url{https://doi.org/10.3389/fams.2022.806549}.

\bibitem[Ferris and Vidal(2012)]{ferris2012perfect}
Andrew~J. Ferris and Guifre Vidal.
\newblock Perfect sampling with unitary tensor networks.
\newblock \emph{Phys. Rev. B}, 85:\penalty0 165146, Apr 2012.
\newblock \doi{10.1103/PhysRevB.85.165146}.
\newblock URL \url{https://link.aps.org/doi/10.1103/PhysRevB.85.165146}.

\bibitem[Gottesman(1997)]{gottesman1997stabilizer}
Daniel Gottesman.
\newblock \emph{Stabilizer codes and quantum error correction. Caltech Ph. D}.
\newblock PhD thesis, Thesis, eprint: quant-ph/9705052, 1997.
\newblock URL \url{https://doi.org/10.48550/arXiv.quant-ph/9705052}.

\bibitem[Gottesman(1998{\natexlab{a}})]{gottesman1998heisenberg}
Daniel Gottesman.
\newblock The heisenberg representation of quantum computers.
\newblock \emph{arXiv quant-ph/9807006}, 1998{\natexlab{a}}.
\newblock URL \url{https://doi.org/10.48550/arXiv.quant-ph/9807006}.

\bibitem[Gottesman(1998{\natexlab{b}})]{gottesman1998theory}
Daniel Gottesman.
\newblock Theory of fault-tolerant quantum computation.
\newblock \emph{Phys. Rev. A}, 57:\penalty0 127--137, Jan 1998{\natexlab{b}}.
\newblock \doi{10.1103/PhysRevA.57.127}.
\newblock URL \url{https://link.aps.org/doi/10.1103/PhysRevA.57.127}.

\bibitem[Hahn et~al.(2022)Hahn, Ferraro, Hultquist, Ferrini, and
  Garc\'{\i}a-\'Alvarez]{hahn2022quantifying}
Oliver Hahn, Alessandro Ferraro, Lina Hultquist, Giulia Ferrini, and Laura
  Garc\'{\i}a-\'Alvarez.
\newblock Quantifying qubit magic resource with gottesman-kitaev-preskill
  encoding.
\newblock \emph{Phys. Rev. Lett.}, 128:\penalty0 210502, May 2022.
\newblock \doi{10.1103/PhysRevLett.128.210502}.
\newblock URL \url{https://link.aps.org/doi/10.1103/PhysRevLett.128.210502}.

\bibitem[Hahn et~al.(2023)Hahn, Ferraro, Hultquist, Ferrini, and
  Garc\'{\i}a-\'Alvarez]{hahn2023erratum}
Oliver Hahn, Alessandro Ferraro, Lina Hultquist, Giulia Ferrini, and Laura
  Garc\'{\i}a-\'Alvarez.
\newblock Erratum: Quantifying qubit magic resource with
  gottesman-kitaev-preskill encoding [phys. rev. lett. 128, 210502 (2022)].
\newblock \emph{Phys. Rev. Lett.}, 131:\penalty0 049901, Jul 2023.
\newblock \doi{10.1103/PhysRevLett.131.049901}.
\newblock URL \url{https://link.aps.org/doi/10.1103/PhysRevLett.131.049901}.

\bibitem[Haug and Kim(2023)]{haug2022scalable}
Tobias Haug and M.S. Kim.
\newblock Scalable measures of magic resource for quantum computers.
\newblock \emph{PRX Quantum}, 4:\penalty0 010301, Jan 2023.
\newblock \doi{10.1103/PRXQuantum.4.010301}.
\newblock URL \url{https://link.aps.org/doi/10.1103/PRXQuantum.4.010301}.

\bibitem[Haug and Piroli(2023)]{haug2022quantifying}
Tobias Haug and Lorenzo Piroli.
\newblock Quantifying nonstabilizerness of matrix product states.
\newblock \emph{Phys. Rev. B}, 107\penalty0 (3):\penalty0 035148, 2023.
\newblock \doi{10.1103/PhysRevB.107.035148}.
\newblock URL \url{https://link.aps.org/doi/10.1103/PhysRevB.107.035148}.

\bibitem[Haug et~al.(2023)Haug, Lee, and Kim]{haug2023efficient}
Tobias Haug, Soovin Lee, and MS~Kim.
\newblock Efficient stabilizer entropies for quantum computers.
\newblock \emph{arXiv:2305.19152}, 2023.
\newblock URL \url{https://doi.org/10.48550/arXiv.2305.19152}.

\bibitem[Heimendahl et~al.(2022)Heimendahl, Heinrich, and
  Gross]{heimendahl2022axiomatic}
Arne Heimendahl, Markus Heinrich, and David Gross.
\newblock The axiomatic and the operational approaches to resource theories of
  magic do not coincide.
\newblock \emph{J. Math. Phys.}, 63\penalty0 (11):\penalty0 112201, 2022.
\newblock \doi{10.1063/5.0085774}.
\newblock URL \url{https://aip.scitation.org/doi/full/10.1063/5.0085774}.

\bibitem[Howard and Campbell(2017)]{howard2017application}
Mark Howard and Earl Campbell.
\newblock Application of a resource theory for magic states to fault-tolerant
  quantum computing.
\newblock \emph{Phys. Rev. Lett.}, 118:\penalty0 090501, Mar 2017.
\newblock \doi{10.1103/PhysRevLett.118.090501}.
\newblock URL \url{https://link.aps.org/doi/10.1103/PhysRevLett.118.090501}.

\bibitem[Jiang and Wang(2023)]{jiang2021lower}
Jiaqing Jiang and Xin Wang.
\newblock Lower bound for the t count via unitary stabilizer nullity.
\newblock \emph{Physical Review Applied}, 19\penalty0 (3):\penalty0 034052,
  2023.
\newblock \doi{10.1103/PhysRevApplied.19.034052}.
\newblock URL
  \url{https://journals.aps.org/prapplied/abstract/10.1103/PhysRevApplied.19.034052}.

\bibitem[Kitaev(2003)]{kitaev2003fault}
A~Yu Kitaev.
\newblock Fault-tolerant quantum computation by anyons.
\newblock \emph{Ann. Phys.}, 303\penalty0 (1):\penalty0 2--30, 2003.
\newblock \doi{10.1016/S0003-4916(02)00018-0}.
\newblock URL
  \url{http://www.sciencedirect.com/science/article/pii/S0003491602000180}.

\bibitem[Lami and Collura(2023)]{lami2023quantum}
Guglielmo Lami and Mario Collura.
\newblock Quantum magic via perfect sampling of matrix product states.
\newblock \emph{arXiv:2303.05536}, 2023.
\newblock URL \url{https://doi.org/10.48550/arXiv.2303.05536}.

\bibitem[Leone et~al.(2022)Leone, Oliviero, and Hamma]{leone2022stabilizer}
Lorenzo Leone, Salvatore F.~E. Oliviero, and Alioscia Hamma.
\newblock Stabilizer r\'enyi entropy.
\newblock \emph{Phys. Rev. Lett.}, 128:\penalty0 050402, Feb 2022.
\newblock \doi{10.1103/PhysRevLett.128.050402}.
\newblock URL \url{https://link.aps.org/doi/10.1103/PhysRevLett.128.050402}.

\bibitem[Leone et~al.(2023{\natexlab{a}})Leone, Oliviero, Esposito, and
  Hamma]{leone2023phase}
Lorenzo Leone, Salvatore F.~E. Oliviero, Gianluca Esposito, and Alioscia Hamma.
\newblock Phase transition in stabilizer entropy and efficient purity
  estimation.
\newblock \emph{arXiv:2302.07895}, 2023{\natexlab{a}}.
\newblock URL \url{https://doi.org/10.48550/arXiv.2302.07895}.

\bibitem[Leone et~al.(2023{\natexlab{b}})Leone, Oliviero, and
  Hamma]{leone2023nonstabilizerness}
Lorenzo Leone, Salvatore F.~E. Oliviero, and Alioscia Hamma.
\newblock Nonstabilizerness determining the hardness of direct fidelity
  estimation.
\newblock \emph{Phys. Rev. A}, 107:\penalty0 022429, Feb 2023{\natexlab{b}}.
\newblock \doi{10.1103/PhysRevA.107.022429}.
\newblock URL \url{https://link.aps.org/doi/10.1103/PhysRevA.107.022429}.

\bibitem[Liu and Winter(2022)]{liu2022many}
Zi-Wen Liu and Andreas Winter.
\newblock Many-body quantum magic.
\newblock \emph{PRX Quantum}, 3:\penalty0 020333, May 2022.
\newblock \doi{10.1103/PRXQuantum.3.020333}.
\newblock URL \url{https://link.aps.org/doi/10.1103/PRXQuantum.3.020333}.

\bibitem[Nielsen and Chuang(2011)]{nielsen2011quantum}
Michael~A. Nielsen and Isaac~L. Chuang.
\newblock \emph{Quantum Computation and Quantum Information: 10th Anniversary
  Edition}.
\newblock Cambridge University Press, 2011.
\newblock ISBN 9781107002173.
\newblock \doi{doi:10.1017/CBO9780511976667}.

\bibitem[Odavi{\'c} et~al.(2022)Odavi{\'c}, Haug, Torre, Hamma, Franchini, and
  Giampaolo]{odavic2022complexity}
J~Odavi{\'c}, T~Haug, G~Torre, A~Hamma, F~Franchini, and SM~Giampaolo.
\newblock Complexity of frustration: a new source of non-local
  non-stabilizerness.
\newblock \emph{arXiv:2209.10541}, 2022.
\newblock URL \url{https://doi.org/10.48550/arXiv.2209.10541}.

\bibitem[Oliviero et~al.(2022{\natexlab{a}})Oliviero, Leone, and
  Hamma]{oliviero2022magic}
Salvatore F.~E. Oliviero, Lorenzo Leone, and Alioscia Hamma.
\newblock Magic-state resource theory for the ground state of the
  transverse-field ising model.
\newblock \emph{Phys. Rev. A}, 106:\penalty0 042426, Oct 2022{\natexlab{a}}.
\newblock \doi{10.1103/PhysRevA.106.042426}.
\newblock URL \url{https://link.aps.org/doi/10.1103/PhysRevA.106.042426}.

\bibitem[Oliviero et~al.(2022{\natexlab{b}})Oliviero, Leone, Hamma, and
  Lloyd]{oliviero2022measuring}
Salvatore F.~E. Oliviero, Lorenzo Leone, Alioscia Hamma, and Seth Lloyd.
\newblock Measuring magic on a quantum processor.
\newblock \emph{npj Quantum Information}, 8\penalty0 (1):\penalty0 148,
  2022{\natexlab{b}}.
\newblock \doi{10.1038/s41534-022-00666-5}.
\newblock URL \url{https://www.nature.com/articles/s41534-022-00666-5}.

\bibitem[Oliviero et~al.(2022{\natexlab{c}})Oliviero, Leone, Lloyd, and
  Hamma]{oliviero2022black}
Salvatore~FE Oliviero, Lorenzo Leone, Seth Lloyd, and Alioscia Hamma.
\newblock Black hole complexity, unscrambling, and stabilizer thermal machines.
\newblock \emph{arXiv:2212.11337}, 2022{\natexlab{c}}.
\newblock URL \url{https://doi.org/10.48550/arXiv.2212.11337}.

\bibitem[Perez-Garcia et~al.(2007)Perez-Garcia, Verstraete, Wolf, and
  Cirac]{perez2007matrix}
D~Perez-Garcia, F~Verstraete, MM~Wolf, and JI~Cirac.
\newblock Matrix product state representations.
\newblock \emph{Quantum Inf. Comp.}, 7\penalty0 (5):\penalty0 401--430, 2007.
\newblock \doi{10.26421/QIC7.5-6-1}.
\newblock URL \url{https://arxiv.org/abs/quant-ph/0608197}.

\bibitem[Preskill(1998)]{preskill1998fault}
John Preskill.
\newblock Fault-tolerant quantum computation.
\newblock In \emph{Introduction to quantum computation and information}, pages
  213--269. World Scientific, 1998.
\newblock \doi{10.1142/9789812385253_0008}.
\newblock URL
  \url{https://www.worldscientific.com/doi/10.1142/9789812385253_0008}.

\bibitem[Regula(2017)]{regula2017convex}
Bartosz Regula.
\newblock Convex geometry of quantum resource quantification.
\newblock \emph{J. Phys. A: Math. Theor.}, 51\penalty0 (4):\penalty0 045303,
  2017.
\newblock \doi{10.1088/1751-8121/aa9100}.
\newblock URL
  \url{https://iopscience.iop.org/article/10.1088/1751-8121/aa9100}.

\bibitem[Sarkar et~al.(2020)Sarkar, Mukhopadhyay, and
  Bayat]{sarkar2020characterization}
Saubhik Sarkar, Chiranjib Mukhopadhyay, and Abolfazl Bayat.
\newblock Characterization of an operational quantum resource in a critical
  many-body system.
\newblock \emph{New J. Phys.}, 22\penalty0 (8):\penalty0 083077, 2020.
\newblock \doi{10.1088/1367-2630/aba919}.
\newblock URL
  \url{https://iopscience.iop.org/article/10.1088/1367-2630/aba919}.

\bibitem[Seddon et~al.(2021)Seddon, Regula, Pashayan, Ouyang, and
  Campbell]{seddon2021quantifying}
James~R. Seddon, Bartosz Regula, Hakop Pashayan, Yingkai Ouyang, and Earl~T.
  Campbell.
\newblock Quantifying quantum speedups: Improved classical simulation from
  tighter magic monotones.
\newblock \emph{PRX Quantum}, 2:\penalty0 010345, Mar 2021.
\newblock \doi{10.1103/PRXQuantum.2.010345}.
\newblock URL \url{https://link.aps.org/doi/10.1103/PRXQuantum.2.010345}.

\bibitem[Sewell and White(2022)]{sewell2022mana}
Troy~J Sewell and Christopher~David White.
\newblock Mana and thermalization: Probing the feasibility of near-clifford
  hamiltonian simulation.
\newblock \emph{Phys. Rev. B}, 106\penalty0 (12):\penalty0 125130, 2022.
\newblock \doi{10.1103/PhysRevB.106.125130}.
\newblock URL
  \url{https://journals.aps.org/prb/abstract/10.1103/PhysRevB.106.125130}.

\bibitem[Shor(1996)]{shor1996fault}
Peter~W Shor.
\newblock Fault-tolerant quantum computation.
\newblock In \emph{Proceedings of 37th conference on foundations of computer
  science}, pages 56--65. IEEE, 1996.
\newblock \doi{10.1109/SFCS.1996.548464}.
\newblock URL \url{https://ieeexplore.ieee.org/document/548464}.

\bibitem[Silvi et~al.(2019)Silvi, Tschirsich, Gerster, Jünemann, Jaschke,
  Rizzi, and Montangero]{silvi2019tensor}
Pietro Silvi, Ferdinand Tschirsich, Matthias Gerster, Johannes Jünemann,
  Daniel Jaschke, Matteo Rizzi, and Simone Montangero.
\newblock The tensor networks anthology: Simulation techniques for many-body
  quantum lattice systems.
\newblock \emph{SciPost Phys. Lect. Notes}, page~8, 2019.
\newblock \doi{10.21468/SciPostPhysLectNotes.8}.
\newblock URL \url{https://scipost.org/10.21468/SciPostPhysLectNotes.8}.

\bibitem[Tirrito et~al.(2023)Tirrito, Tarabunga, Lami, Chanda, Leone, Oliviero,
  Dalmonte, Collura, and Hamma]{tirrito2023quantifying}
Emanuele Tirrito, Poetri~Sonya Tarabunga, Gugliemo Lami, Titas Chanda, Lorenzo
  Leone, Salvatore~FE Oliviero, Marcello Dalmonte, Mario Collura, and Alioscia
  Hamma.
\newblock Quantifying non-stabilizerness through entanglement spectrum
  flatness.
\newblock \emph{arXiv:2304.01175}, 2023.
\newblock URL \url{https://doi.org/10.48550/arXiv.2304.01175}.

\bibitem[Turkeshi et~al.(2023)Turkeshi, Schir{\`o}, and
  Sierant]{turkeshi2023measuring}
Xhek Turkeshi, Marco Schir{\`o}, and Piotr Sierant.
\newblock Measuring magic via multifractal flatness.
\newblock \emph{arXiv:2305.11797}, 2023.
\newblock URL \url{https://doi.org/10.48550/arXiv.2305.11797}.

\bibitem[Veitch et~al.(2014)Veitch, Mousavian, Gottesman, and
  Emerson]{veitch2014resource}
Victor Veitch, SA~Hamed Mousavian, Daniel Gottesman, and Joseph Emerson.
\newblock The resource theory of stabilizer quantum computation.
\newblock \emph{New J. Phys.}, 16\penalty0 (1):\penalty0 013009, 2014.
\newblock \doi{10.1088/1367-2630/16/1/013009}.
\newblock URL \url{https://doi.org/10.1088/1367-2630/16/1/013009}.

\bibitem[Wang et~al.(2019)Wang, Wilde, and Su]{wang2019quantifying}
Xin Wang, Mark~M Wilde, and Yuan Su.
\newblock Quantifying the magic of quantum channels.
\newblock \emph{New J. Phys.}, 21\penalty0 (10):\penalty0 103002, 2019.
\newblock \doi{10.1088/1367-2630/ab451d}.
\newblock URL \url{https://doi.org/10.1088/1367-2630/ab451d}.

\bibitem[White et~al.(2021)White, Cao, and Swingle]{white2021conformal}
Christopher~David White, ChunJun Cao, and Brian Swingle.
\newblock Conformal field theories are magical.
\newblock \emph{Phys. Rev. B}, 103:\penalty0 075145, Feb 2021.
\newblock \doi{10.1103/PhysRevB.103.075145}.
\newblock URL \url{https://link.aps.org/doi/10.1103/PhysRevB.103.075145}.

\end{thebibliography}

\appendix

\section{Details on the violation of monotonicity}
\label{sec:details_counter}

In this appendix, we provide further details on the counterexamples presented in Sec.~\ref{sec:counter_example}. First, we explain how we arrived at the state~\eqref{eq:counter_example}.

We initially looked for violations of the strong monotonicity condition~\eqref{eq:strong_monotonicity}. To this end, we defined the functional
\begin{equation}\label{eq:functional}
\Delta_n(\ket{\psi})=M_n(\ket{\psi})-\sum_{a=0}^1 p_{a}M_n(\ket{\psi_a})\,,
\end{equation}
where
\begin{equation}
\ket{\psi_a}=\frac{1}{\sqrt{p_a}}\Pi_a \ket{\psi}\,,
\end{equation}
and $\Pi_a=|a\rangle\langle a|\otimes \openone_{2,3,\ldots N}$, while $p_{a}=\braket{\psi|\Pi_a|\psi}$. Namely, $\ket{\psi_a}$ are the states obtained after a computational-basis measurement of qubit $1$, and $p_{a}$ are the corresponding probabilities. Note that a negative value of $\Delta_n(\ket{\psi})$ implies a violation of the strong monotonicity condition. 

Next, using a gradient-descent method, we have looked numerically for the minimum of~\eqref{eq:functional} for different values of $N$. For $N\leq 3$, we found $\Delta_n(\ket{\psi})\geq 0$, while for $N=4$ we found that there are states for which $\Delta_n(\ket{\psi})<0$ for $0\leq n < 2$. 

The state~\eqref{eq:counter_example} was obtained as the result of this minimization procedure for $N=4$ (decimal numbers were rounded as fractions, e.g. $0.408248\simeq 1/\sqrt{6}$, by guesswork). Remarkably, by analytical inspection, we found that this state also has additional properties. Most notably, it is an eigenstate of the Clifford operator ~\eqref{eq:unitary}. This made it possible to devise the Clifford protocol presented in Sec.~\ref{sec:violation_mono}, which was done by analytic inspection and guesswork. Therefore, although the state was found by looking for violations of the strong-monotonicity condition, it turned out to also provide a counterexample to the monotonicity condition~\eqref{eq:monotonicity_condition}.

The same method applied to $N=5$ qubits becomes cumbersome. While we were able to carry out the minimization procedure numerically, the states which were obtained in this way did not have the desirable properties of~\eqref{eq:counter_example}. In particular, using this method we did not find states which were eigenstates of Clifford operators.

Next, we provide further details on the state $|\psi^\varepsilon\rangle$ of~\eqref{eq:psi_eps_state}. We show in particular that its SE for $n>1$ is bounded, as anticipated in the main text. 

We start by defining
\begin{equation}
	G_n(|\psi^\varepsilon\rangle) = \sum_{P\in \mathcal{P}_N} \frac{\langle \psi^\varepsilon| P|\psi^\varepsilon\rangle^{2n}}{2^{N}}\,,
\end{equation}
so that $M_n(|\psi^\varepsilon\rangle)=\log G_n(|\psi^\varepsilon\rangle)/(1-n)$. 
Next, we note that the state $\ket{\psi^\varepsilon}$ is invariant under permutation of qubits. Therefore, we can simplify
\begin{widetext}
\begin{equation}\label{eq:function_f}
	G_n(|\psi^\varepsilon\rangle) = \frac{1}{2^N}\sum_{N_z=0}^N \sum_{N_x=0}^{N-N_z} \sum_{N_y=0}^{N-N_z-N_x} \binom{N}{N_z, N_x, N_y,N-N_z-N_x-N_y}\langle \psi^\varepsilon| \prod_{i=1}^{N_z} \sigma^z_i \prod_{j=N_z+1}^{N_z+N_x} \sigma^x_{j} \prod_{k=N_z+N_x+1}^{N_z+N_x+N_y} \sigma^y_{k}|\psi^\varepsilon \rangle^{2n} \,.
\end{equation}
In addition, the expectation value of Pauli matrices can be evaluated explicitly, yielding
\begin{align}
\langle \psi^\varepsilon|\prod_{i=1}^{N_z} \sigma^z_i \prod_{j=N_z+1}^{N_z+N_x} \sigma^x_{j} \prod_{k=N_z+N_x+1}^{N_z+N_x+N_y} \sigma^y_{k}|\psi^\varepsilon \rangle&=\frac{\delta_{N_x,0} \delta_{N_y,0} +\varepsilon^2 3^{-(N_x+N_y+N_z)/2}}{\mathcal{N}_\varepsilon}\nonumber\\
+&\frac{2\varepsilon\sin(\beta)^{N_x+N_y}\cos(\beta)^{N-N_x-N_y}\cos[\pi(N-N_x+N_y)/4]}{\mathcal{N}_\varepsilon}\,.\label{eq:form_factor}
\end{align}
\end{widetext}
The number of terms in the sum of Eq.~\eqref{eq:function_f} scales polynomially in $N$, and can be evaluated in a few seconds for up to $N\sim 50$ qubits. We have done this for different values of $n>1$, and confirmed that $M_n(\ket{\psi^\varepsilon})$ is bounded in $N$. In fact, it is not difficult, although a bit cumbersome, to prove this explicitly, as we now sketch. 

\begin{figure}
	\includegraphics[scale=0.52]{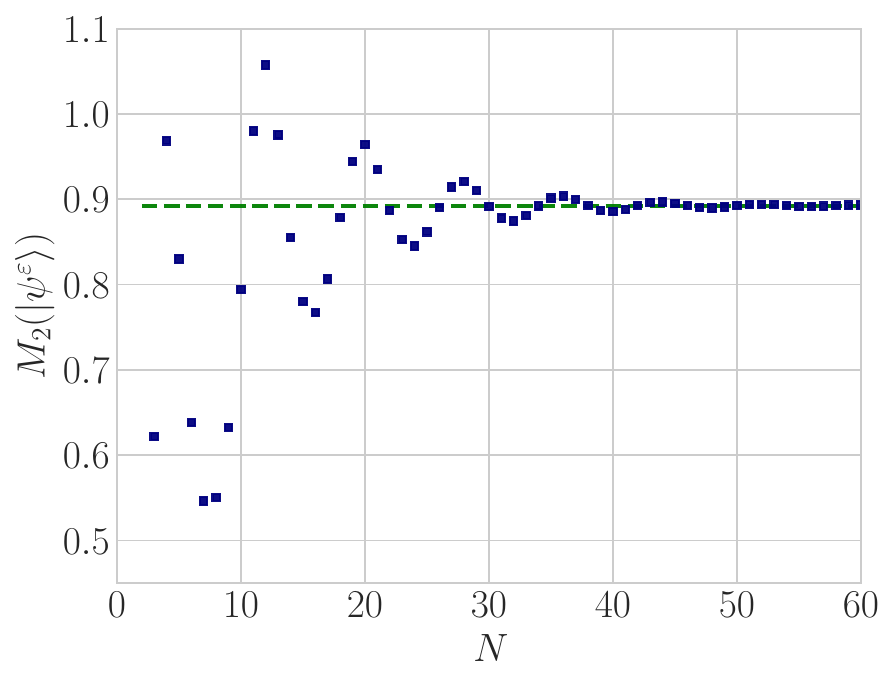}
	\caption{Numerical test of the asymptotic prediction~\eqref{eq:final_bound} (for $n=2$). Symbols correspond to the exact evaluation of~\eqref{eq:function_f}, while the dashed line is the prediction~\eqref{eq:final_bound}. Here we chose $\varepsilon=0.5$.}
	\label{fig:numerical_test}
\end{figure}

We can write $G_n(\ket{\psi^\varepsilon})\simeq ({\rm I})/\mathcal{N}_\varepsilon^{2n}+({\rm II})/\mathcal{N}_\varepsilon^{2n}$, where
\begin{align}
({\rm I})&=  \frac{1}{2^N}\sum_{N_z\geq 0}\binom{N}{N_z}\left[1+\varepsilon^2 3^{-N_z/2}\right.\nonumber\\
&+\left.2\varepsilon\cos(\beta)^{N}\cos(\pi N/4)\right]^{2n}\,,
\end{align}
and
\begin{align}
    ({\rm II})&= \frac{1}{2^N}\sum_{\substack{N_z\geq 0\\ N_x\|N_y\neq 0}}\binom{N}{N_z, N_x, N_y,N-N_x-N_y-N_z} \nonumber\\
    &\times (\varepsilon^2 3^{-(N_x+N_y+N_z)/2}+ 2\varepsilon \sin(\beta)^{N_x+N_y}\nonumber\\
    &\times \cos(\beta)^{N-N_x-N_y}\cos[\pi(N-N_x+N_y)/4] )^{2n}\,.
\end{align}
First, we note that
\begin{equation}
    ({\rm I})\geq  \frac{1}{2^N}\sum_{N_z\geq 0}\binom{N}{N_z}(1-\delta_N)^{2n}=  (1-\delta_N)^{2n}\,,
\end{equation}
where $\delta_N=|2\varepsilon\cos(\beta)^{N}\cos(\pi N/4)|$ is an exponentially vanishing term. Next, we show that $(\rm II)$ is exponentially small. Indeed, using the triangular inequality for the absolute value, and the convexity inequality $(|A_1+A_2|)^{2n}\leq 2^{2n-1}(|A_1|^{2n}+|A_2|^{2n}|)$, we obtain 
\begin{align}
    ({\rm II})&\leq \frac{2^{2n-1}}{2^N}\sum_{N_x,N_y,N_z}\binom{N}{N_z, N_x, N_y,N-N_x-N_y-N_z} \nonumber\\
    &\times \{(\varepsilon^2 3^{-(N_x+N_y+N_z)/2})^{2n}+ \nonumber\\
    &+[2\varepsilon \sin(\beta)^{N_x+N_y}\
    \cos(\beta)^{N-N_x-N_y}]^{2n}\}\,,
\end{align}
where we have extended the sum to $N_x,N_y,N_z\geq 0$ such that $N_x+N_y+N_x=N$. Applying the multinomial theorem, we obtain 
\begin{align}
    |({\rm II})|&\leq  2^{2n-1}\frac{\varepsilon^{4n}}{2^{N}} \left(1+3 \left(\frac{1}{\sqrt{3}}\right)^{2n} \right)^{N}\nonumber\\
    &+2^{2n-1}\frac{(2\varepsilon)^{2n}}{2^N}(2\cos(\beta)^{2n}+2\sin(\beta)^{2n})^N\,.
\end{align}
For $n>1$ we see that $({\rm II})$ is exponentially small in $N$. Therefore, using $\log\mathcal{N}_\varepsilon\simeq \log(1+\varepsilon^2)$ (up to exponentially small terms in $N$) we have
\begin{equation}\label{eq:final_bound}
    M_{n}(\ket{\psi^{\varepsilon}})\leq \frac{ 2n}{n-1}\log(1+\varepsilon^2)+\tilde\delta_N\,,
\end{equation}
where $\tilde\delta_N$ is an exponentially small correction. We have verified~\eqref{eq:final_bound}, based on exact evaluation of~\eqref{eq:function_f}. We report an example of our numerical data in Fig.~\ref{fig:numerical_test}, showing excellent agreement. 

\section{Additional numerical results}\label{sec:numerics_appendix}

\begin{figure}[htbp]
	\centering	
	\subfigimg[width=0.3\textwidth]{}{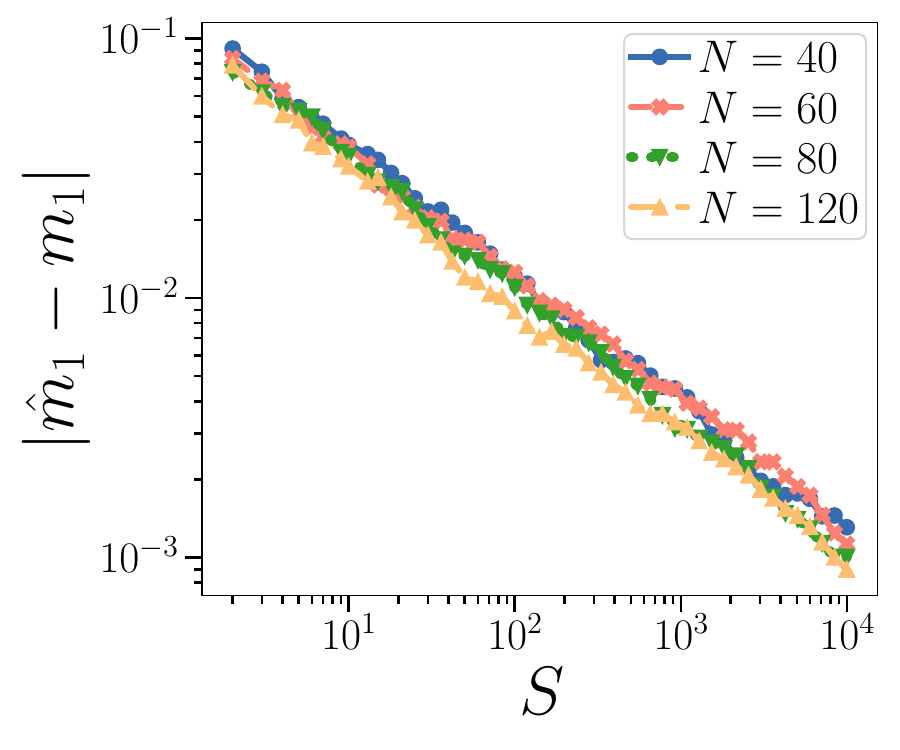}
	\caption{Average error in estimation of the von Neumann SE density $\vert \hat{m}_1 -m_1\vert$ against number of samples $S$. $\hat{m}_1$ is the estimated SE density, while $m_1$ is a highly accurate estimation from $S=10^5$ samples. The error is averaged over 100 random instances.  We show the ground state of the Heisenberg model for half-filling, $\Delta=0.95$ and different qubit number $N$.  
	}
	\label{fig:sample}
\end{figure}

We provide further details about the numerical results discussed in Sec.~\ref{sec:numerics}. First, we present additional numerical data about the perfect-MPS sampling method for the von Neumann SE. In Fig.~\ref{fig:sample} we show how the approximation error decreases with the number of samples $S$. We plot the average difference $|\hat{m}_1-m_1|$, where $\hat{m}_1$ is an estimate for $m_1$, computed by averaging over a given number of samples $S$, while $m_1$ is our best prediction, computed by averaging over a very large value of samples, $S_{\rm max}$ (in this case, $S_{\rm max} = 10^5$). The plot clearly shows the scaling $|\hat{m}_1-m_1| \sim S^{-1/2}$, as expected.

\begin{figure*}[htbp]
	\centering	
	\subfigimg[width=0.3\textwidth]{a}{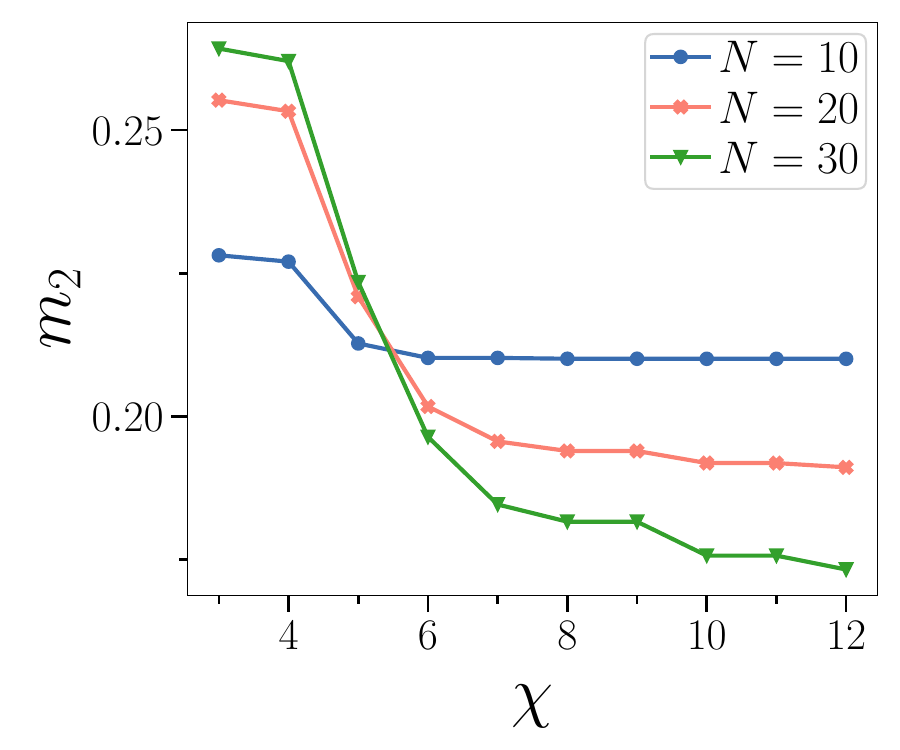}
	\subfigimg[width=0.3\textwidth]{b}{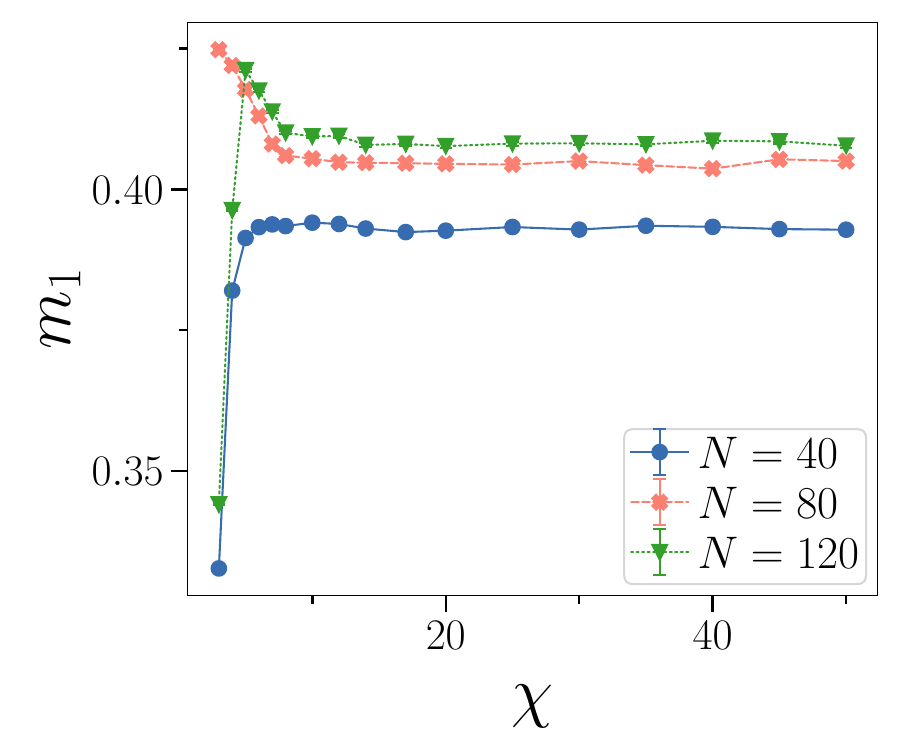}
 	\subfigimg[width=0.3\textwidth]{c}{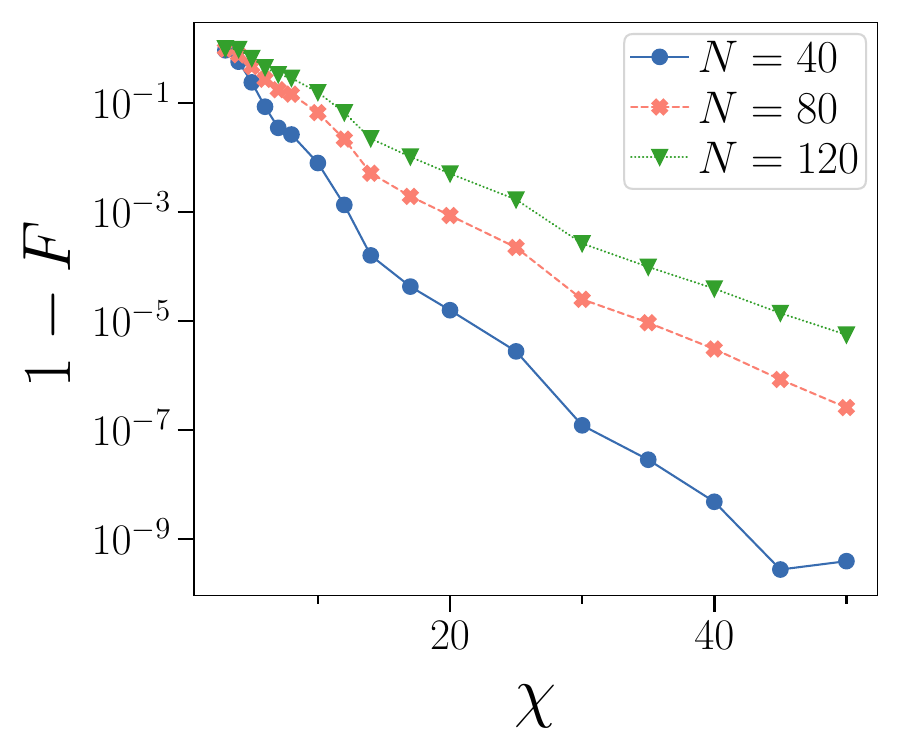}
	\caption{\idg{a} R\'enyi-$2$ SE density $m_2$ against bond dimension $\chi$. The plot corresponds to the MPS representing the ground state of the Heisenberg model at half-filling. Here, we have used the method of Ref.~\cite{haug2022quantifying}.
    \idg{b} Estimated value of $m_1$ as a function of $\chi$, using the method presented in Sec.~\ref{sec:numerics}
    \idg{c} Fidelity $F=\vert \braket{\Psi_N(\chi)\vert {{\rm GS}}}\vert^2$ between the MPS approximation for the ground state, $\ket{\Psi_N(\chi)}$, and the true ground state $\ket{{\rm GS}}$ as function of $\chi$. The true ground state $\ket{{\rm GS}}$ is estimated by taking a very large bond dimension (in this case, $\chi \simeq 200$). In all plots, we set $\Delta=0.95$.
	}
	\label{fig:bond}
\end{figure*}

Next, we study the dependence of the estimated values of $m_1$ and $m_2$ with the bond dimensions for the ground-state of the Heisenberg model at half-filling. Our data are reported in Fig.~\ref{fig:bond}. In Fig.~\ref{fig:bond}a, we show the R\'enyi-$2$ SE density $M_2/N=m_2$ as function of bond dimension $\chi$, which was computed using the method of Ref.~\cite{haug2022quantifying}. We find that the value of $m_2$ decreases and converges to a constant for increasing $\chi$. In Fig.~\ref{fig:bond}b, we show a similar plot for the SE, computed with the method presented in Sec.~\ref{sec:numerics}. Once again, we see convergence as the bond dimension is increased. Finally, in Fig.~\ref{fig:bond}c, we show the fidelity $F=\vert \braket{\Psi_N(\chi)\vert {{\rm GS}}}\vert^2$ between the MPS approximation with bond dimension $\chi$, $\ket{\Psi_N(\chi)}$, and the true ground state $\ket{{\rm GS}}$. The true ground-state $\ket{{\rm GS}}$ is estimated by taking a very large bond dimension (in this case, $\chi \simeq 200$). 

\end{document}